\documentclass[preprint,12pt,authoryear]{elsarticle}

\usepackage{xcolor}
\usepackage{amssymb}
\usepackage{amsmath}
\usepackage{subcaption}
\usepackage{mathtools}
\usepackage{stmaryrd}
\usepackage{xspace}
\usepackage{colonequals}
\usepackage{hyperref}

\newcommand{\myvec}[1]{#1}
\def\newnot#1{\label{#1}}

\newcommand{\fenics}{{\mbox{FEniCS}}\xspace}
\newcommand{\ponedgptwo}{\mbox{P1$_\textrm{DG}$-P2}\xspace}
\newcommand{\dx}{\,\mathrm{d}x}
\newcommand{\dt}{\,\mathrm{d}t}

\newcommand{\da}{\mbox{{dolfin-adjoint}}\xspace}

\journal{Computer Methods in Applied Mechanics and Engineering}

\begin{document}

\begin{frontmatter}

    \title{Reconstructing wave profiles from inundation data}
\cortext[cor1]{Corresponding author}

\address[simula]{Biomedical Computing, Simula Research Laboratory, 1364 Fornebu, Norway}
    \address[ox]{Mathematical Institute, University of Oxford, OX2 6GG, UK}
\address[ic]{Department of Earth Science and Engineering, Imperial College London, London, SW7 2AZ, UK}

\author[simula]{S.W. Funke\corref{cor1}}
\ead{simon@simula.no}
\author[ox,simula]{P.E. Farrell}
\author[ic]{M.D. Piggott}

\begin{abstract}
This paper applies variational data assimilation to inundation problems
governed by the shallow water equations with wetting and drying.  The
objective of the assimilation is to recover an unknown time-varying wave profile at an open
ocean boundary from inundation observations. This problem is solved with 
derivative-based optimisation and an adjoint wetting and drying
scheme to efficiently compute sensitivity information. The capabilities of this approach are
demonstrated on an idealised sloping beach setup in which the profile of an
incoming wave is reconstructed from wet/dry interface observations. The method is
robust against noisy observations if a regularisation term is added to the optimisation objective.
Finally, the method is applied to a laboratory experiment of the Hokkaido-Nansei-Oki tsunami,
where the wave profile is reconstructed with an error of less than $1$\% of the reference wave signal.
\end{abstract}


\end{frontmatter}



\section{Introduction}
Wetting and drying plays an important role in coastal research for the study of tsunamis~\citep{kowalik2004}, storm surge hazards~\citep{westerink2008}, tidal flats and river estuaries~\citep{zhang2009, xue2010, karna2011}, and other flooding events~\citep{song2010}.
Many algorithms have been proposed for the simulation of wetting and drying processes, both for the shallow-water equations~(\citet{medeiros2012} and the references therein) and for the Navier-Stokes equations~\citep{funke2011}.

In addition to the pure simulation of wetting and drying problems, it is often desirable to study the sensitivity of the
result with respect to changes in the input parameters such as initial and boundary conditions.
The key for the efficient computation of these sensitivities is the adjoint approach \citep{errico1997, gunzburger2003}.
In the context of shallow water modelling without wetting and drying, adjoint models have been successfully used in various applications, ranging from
data assimilation \citep{bagchi1994, gejadze2005, chen2009} and parameter identification \citep{ding2004}, to wave and flood control \citep{kawahara1989, sanders1996, sanders2000, asce2006, samizo2012}.
\citet{blaise2012} successfully reconstructed the initial condition for a tsunami simulation from buoy measurements,
but also emphasized the importance of including wetting and drying in the adjoint model as future work.

The main contribution of this paper is the development of an adjoint model for the shallow water equations
with wetting and drying. The adjoint model computes the sensitivity (or gradient) of the wet/dry interface with respect to boundary conditions at a computational cost equivalent to one linearised shallow water solve.
The adjoint model is then used to efficiently solve data assimilation problems with gradient-based optimisation.
The goal of the data assimilation here is to reconstruct the wave height boundary values that lead to an observed wet/dry interface.

\section{Shallow water model with wetting and drying}
\subsection{Continuous formulation}
\begin{figure}[t]
\centering
        \begin{subfigure}[b]{0.48\textwidth}
                \centering
                \includegraphics[width=\textwidth]{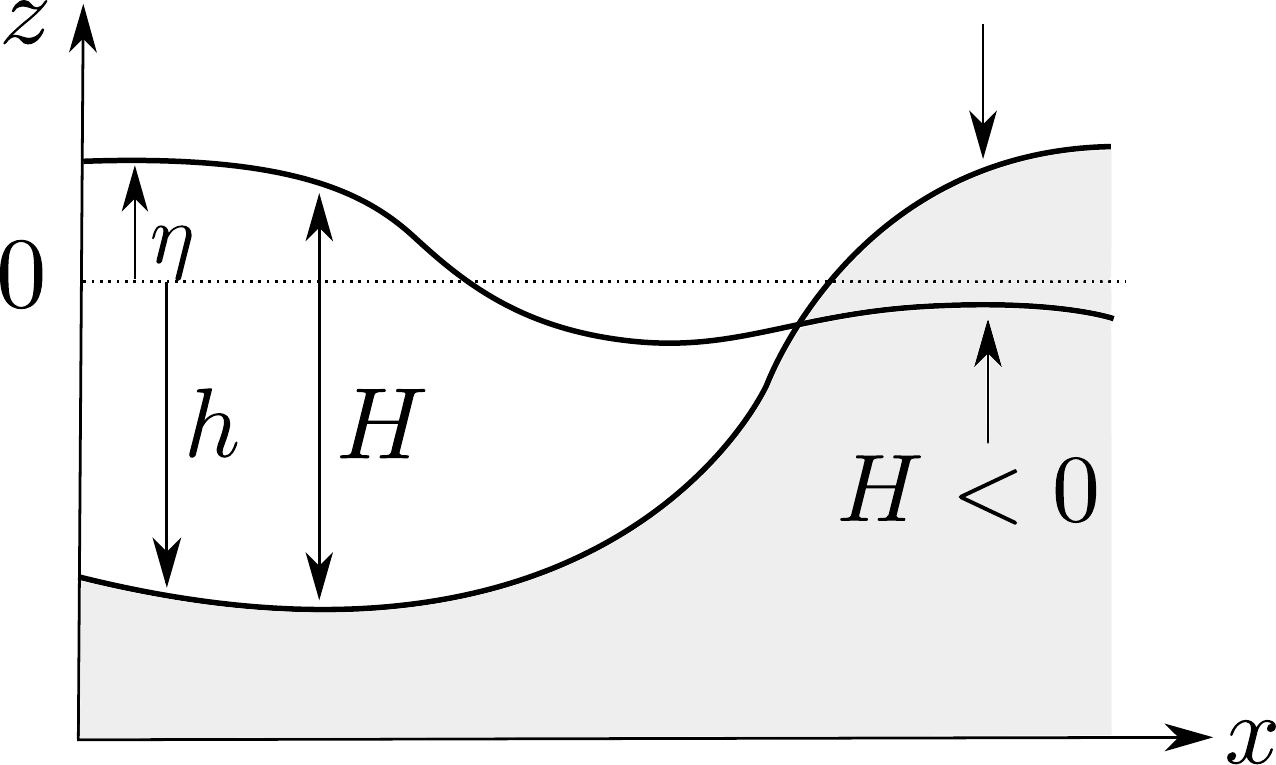}
                \caption{Variables for the static bathymetry (no wetting and drying)\\~}
                \label{fig:wd_optimisation/pics/wd_explained_fixed}
        \end{subfigure}
        ~
        \begin{subfigure}[b]{0.48\textwidth}
                \centering
                \includegraphics[width=\textwidth]{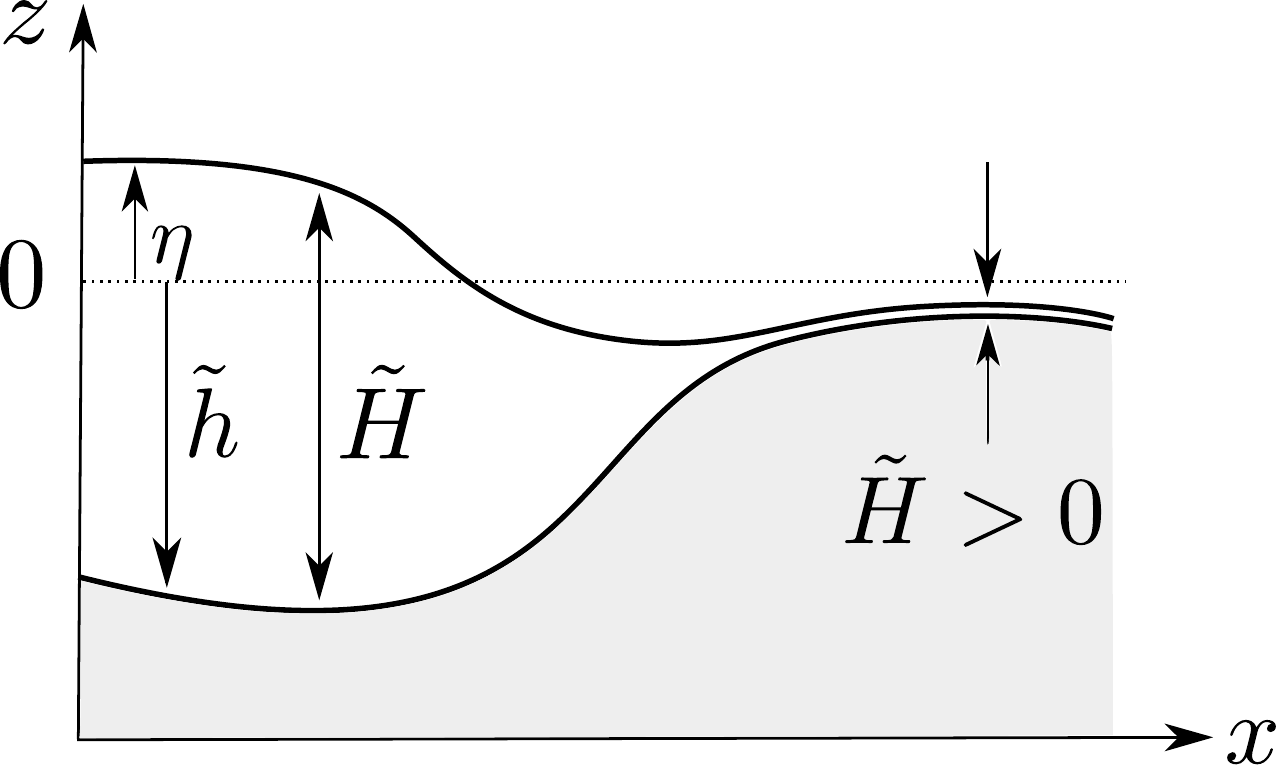}
                \caption{Variables for the bathymetry modified to account for wetting and drying}
                \label{fig:wd_optimisation/pics/wd_explained_variable}
        \end{subfigure}
        \caption{The modified water depth variables for the wetting and drying scheme by \citet{karna2011}.}\label{fig:wd_explained}
\end{figure}

The non-linear shallow water equations with appropriate initial and boundary conditions are considered here in the form
\begin{subequations}
\begin{align}
  \frac{\partial \myvec{u}}{\partial t} + (\myvec u \cdot \nabla) \myvec{u} + g \nabla \eta &= -\frac{c_f(H)}{H} \|\myvec u\|\myvec u && \mbox{in } \Omega \times (0, T), \label{eq:opt_wd_momentum_equation} \\
\frac{\partial \eta}{\partial t} + \nabla \cdot  (H \myvec{u}) & = 0 && \mbox{in }  \Omega \times (0, T), \label{eq:opt_wd_continuitiy_equation} \\
\myvec u \cdot \myvec n & = 0 && \mbox{on } \partial \Omega_{S} \times (0,T), \\
\eta & = \eta_{D} && \mbox{on } \partial \Omega_{D} \times (0,T), \\
\myvec u & = \myvec{u}_0, \quad \eta = \eta_0 && \mbox{at } \Omega \times \{0\},
\end{align}\label{eq:wd_opt_shallow_water_equations}%
\end{subequations}
where $\Omega\subset\mathbb R^2$ is the domain of interest, $T$ is the final time, $\myvec u$ is the unknown depth-averaged velocity, $\eta$ is the unknown free-surface displacement,
$h$ describes the static bathymetry, $H = \eta + h$ is the total water depth, $u_0$ and $\eta_0$ are the initial conditions, and $n$ is the normal vector on the boundary.
The water height variables are sketched in figure~\ref{fig:wd_optimisation/pics/wd_explained_fixed}.
The domain boundary is divided into $\partial \Omega_S$\newnot{partialomegaN}, where a no-normal flow condition is imposed, and $\partial \Omega_D$\newnot{partialomegaD}, where a Dirichlet boundary condition prescribes the free-surface displacement $\eta_D$.
The remaining parameters are the gravitational force $g$ and the friction coefficient in the Ch\'{e}zy-Manning formulation \citep{hervouet2007}
\begin{equation*}
c_f(H) = \frac{g \mu ^ 2}{H^{1/3}},
\end{equation*}
where $\mu$ is the user specified Manning coefficient.

In its standard form, the shallow water equations do not account for wetting
and drying processes. With wetting and drying, the domain
$\Omega$ becomes an unknown variable itself, described by all points where
the total water-level is positive. Hence, equations
\eqref{eq:wd_opt_shallow_water_equations} are extended by the domain equation
\begin{equation}
\Omega(t) = \{(x, y): H(x, y, t) > 0\}.
    \label{eq:wetting_drying_extension}
\end{equation}
The numerical treatment of wetting and drying is challenging, and various
extensions have been proposed and are reviewed in~\citet{medeiros2012}.
A classical approach is to mark individual mesh elements in the computational
domain as wet or dry and remove dry elements from the time step computation.
However, the elemental wet/dry conditions usually involve discontinuous functions,
which complicates the development of the adjoint system.  This can be seen in
the work of \citet{miyaoka2008}, where the wetting and drying algorithm was
ignored in the adjoint computation; instead, the adjoint shallow water
equations without wetting and drying were solved only in the wet area.  Such an
approach cannot provide the sensitivity of the wet/dry interface, which is
needed here for the data assimilation. Therefore, we use an
alternative wetting and drying algorithm proposed by~\citet{karna2011},
motivated by the fact that their numerical scheme is differentiable.

\begin{figure}[p]
\centering
        \begin{subfigure}[b]{0.48\textwidth}
                \centering
                \includegraphics[width=\textwidth]{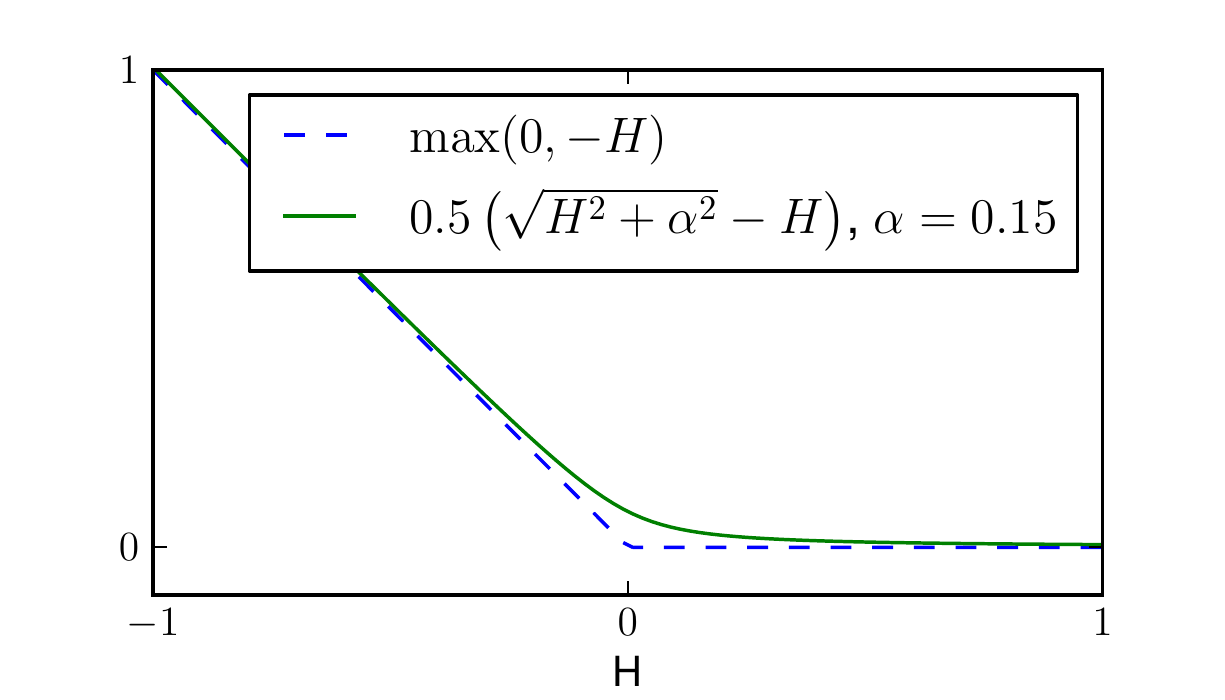}
                \caption{Smooth approximation of $\max(0, -H)$, used to enforce a positive water level in equation~\eqref{eq:wd_opt_tilde_H_def}\\}
                \label{fig:wd_optimisation/pics/plot_f}
        \end{subfigure}
        ~
        \begin{subfigure}[b]{0.48\textwidth}
                \centering
                \includegraphics[width=\textwidth]{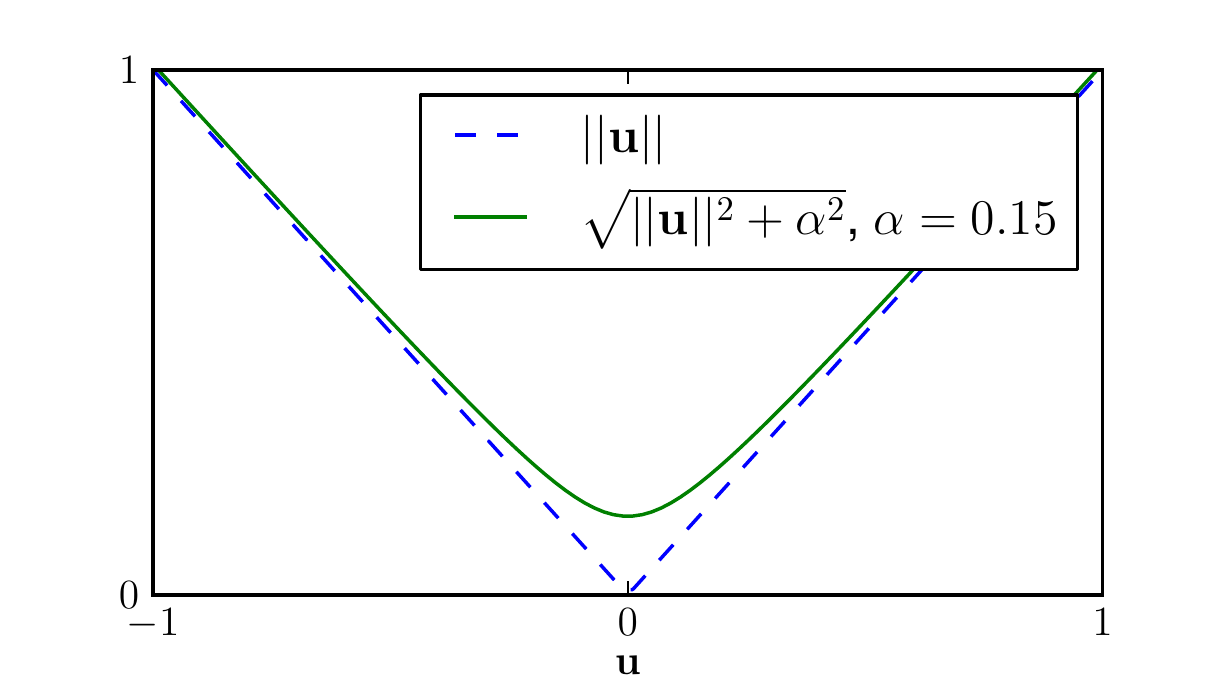}
                \caption{Smooth approximation of the norm, used in the drag term of the momentum equation~\eqref{eq:opt_wd_momentum_equation}\\~}
                \label{fig:wd_optimisation/pics/plot_norm}
        \end{subfigure}
        \\
        ~
        \begin{subfigure}[b]{0.8\textwidth}
                \centering
                \includegraphics[width=\textwidth]{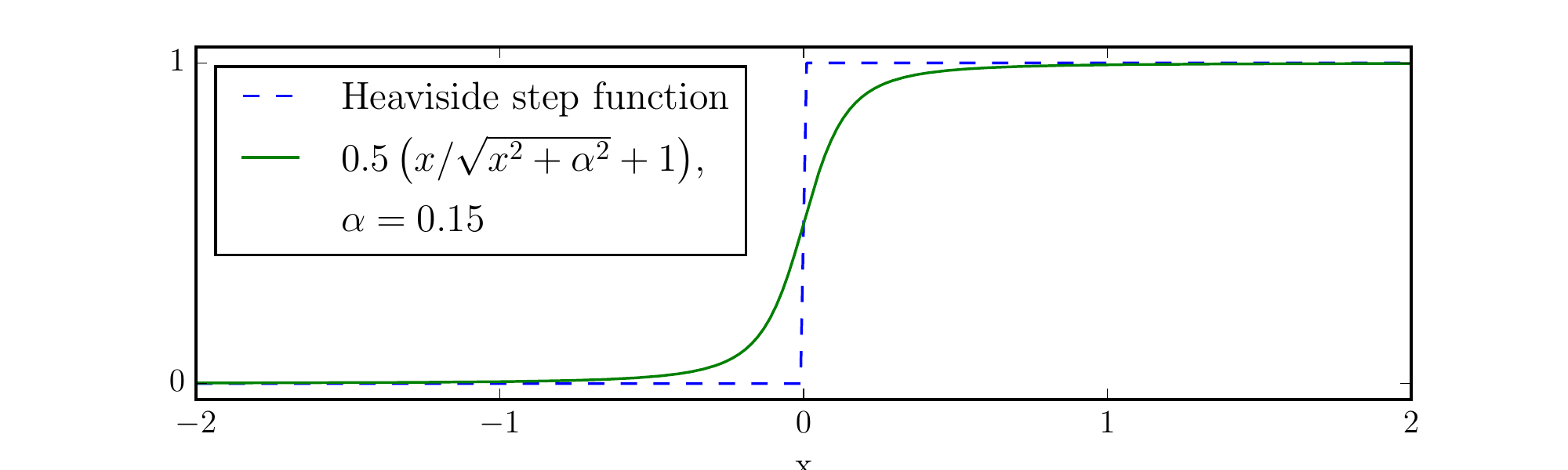}
                \caption{Smooth approximation of the Heaviside step
                function~\eqref{eq:wd_opt_smooth_heaviside_approx}, used in the
            functional of interest as an indicator function for dry areas. Note
        that the smooth representation is equivalent to $f'(x) + 1$, where $f$
    is defined in equation~\eqref{eq:wd_function_f}}
                \label{fig:wd_optimisation/pics/plot_heaviside}
        \end{subfigure}
        \caption{Smooth approximations of the non-differentiable functions that occur in the problem formulation. This paper uses the same smoothness constant $\alpha$ for all approximations.
}\label{fig:wd_smooth_representations}
\end{figure}

The wetting and drying algorithm developed is based on the idea of replacing the static bathymetry $h$ with a dynamic bathymetry $\tilde h$, which moves such that the water level remains always positive.
This dynamic bathymetry is defined as
\begin{equation*}
\tilde h(x,t) \coloneqq h + f(H),
\end{equation*}
where $f$ is a smooth function that ensures the positiveness of the total water depth, that is:
\begin{equation}
\tilde H \coloneqq \eta + \tilde h > 0. \label{eq:wd_opt_tilde_H_def}
\end{equation}
The modified variables for the dynamic bathymetry approach are sketched in figure~\ref{fig:wd_optimisation/pics/wd_explained_variable}.
For the function $f$, \citet{karna2011} suggest a smooth approximation of the maximum operator:
\begin{equation}
f(H) \coloneqq \frac{1}{2} \left( \sqrt{H^2 + \alpha^2} - H \right) \approx \max(0, -H). \label{eq:wd_function_f}
\end{equation}
This function choice, plotted in figure~\ref{fig:wd_optimisation/pics/plot_f}, is also used in this work.
The parameter $\alpha>0$ controls the accuracy of the approximation to the $\max$ operator.
\citet{karna2011} provides a guideline for determining a suitable estimate for this parameter:
\begin{equation}
\alpha \coloneqq d_e \| \nabla h \|, \label{eq:epsilon_definition_in_wd}
\end{equation}
where $d_e$ is a typical length scale of a representative element in the computational mesh.
By ensuring that $\alpha \rightarrow 0$ as the mesh size goes to zero, this ensures
consistency of the discretisation with the nonsmooth equations.

The modified shallow water equations that include wetting and drying are obtained from the original equations~\eqref{eq:wd_opt_shallow_water_equations} by replacing the total depth $H$ with its dynamic variant $\tilde H$ and
including the time derivative of the dynamic bathymetry $\tilde h$ in the continuity equation to account for the temporal variability of the bathymetry:
\begin{equation}
\begin{aligned}
  \frac{\partial \myvec{u}}{\partial t} + (\myvec{u} \cdot \nabla) \myvec{u} + g \nabla \eta &= -\frac{c_f(\tilde H)}{\tilde H} \|\myvec u\|\myvec u && \mbox{in } \Omega \times (0, T), \\
\frac{\partial \eta}{\partial t} + \frac{\partial \tilde h}{\partial t} + \nabla \cdot  (\tilde H \myvec{u}) &= 0 && \mbox{in } \Omega \times (0, T),  \\
\myvec{u} \cdot \myvec n & = 0 && \mbox{on } \partial \Omega_{S} \times (0,T),\\
\eta & = \eta_{D} && \mbox{on }  \partial \Omega_{D} \times (0, T), \\
\myvec{u} & = \myvec{u}_0, \quad \eta = \eta_0 && \mbox{at } t=0.
\end{aligned}\label{eq:modified_shallow_water_with_wetting_and_drying}%
\end{equation}
To avoid non-differentiable functions in the continuous formulation, the norm operator in \eqref{eq:modified_shallow_water_with_wetting_and_drying} is replaced by a smooth approximation:
\begin{equation*}
  \|\myvec u\|\myvec \approx \sqrt{ \|\myvec u\|^2 + \alpha^2 },
\end{equation*}
with the same $\alpha$ constant as above.
A plot of this approximation function is given in figure~\ref{fig:wd_optimisation/pics/plot_norm}.

\subsection{Spatial discretisation}\label{sec:wd_opt_spatial_discretisation}
The modified shallow water equations~\eqref{eq:modified_shallow_water_with_wetting_and_drying} are discretised in space with a mixed continuous-discontinuous finite element method.
A general introduction to discontinuous Galerkin methods can be found in \citet{hesthaven2008}.

The discrete function spaces are constructed with the \ponedgptwo finite element pair \citep{cotter2008b, comblen2008}.
Let $V$ and $W$ denote the associated function spaces for the velocity and free-surface displacement fields, respectively.
The weak formulation is obtained by multiplying the two partial differential equations in~\eqref{eq:modified_shallow_water_with_wetting_and_drying} with test functions $\myvec \Psi \in V$ and $\Phi \in W$ and integrating over the domain $\Omega$. The resulting discretised variational problem is to find $\myvec u \in V, \eta \in W$ such that $\forall~\myvec \Psi \in V,~\Phi \in W$:
\begin{subequations}
\begin{align}
  & \left< \frac{\partial \myvec{u}}{\partial t}, \myvec \Psi\right>_\Omega + \left< (\myvec{u} \cdot \nabla) \myvec{u}, \myvec \Psi \right>_\Omega
- \sum_{e \in E}\left( \left<\{ \myvec u^+ \} \llbracket \myvec u \rrbracket, \myvec \Psi^+ \right>_{e} - \left<\{ \myvec u^- \} \llbracket \myvec u \rrbracket, \myvec \Psi^- \right>_{e} \right)
\notag \\
  & + \left< g \nabla \eta, \myvec \Psi \right>_\Omega - g \left< \eta - \eta_D, \myvec \Psi \cdot \myvec n\right>_{\partial \Omega_{D}}= -\left< \frac{c_f(\tilde H)}{\tilde H} \|\myvec u\|\myvec u, \myvec \Psi \right>_\Omega, \label{eq:weak_modified_shallow_water_with_wetting_and_drying_momentum} \\
& \left< \frac{\partial \tilde H}{\partial t}, \Phi \right>_\Omega - \left< (\tilde H \myvec{u}), \nabla \Phi \right>_\Omega + \left< \tilde H \myvec{u} \cdot \myvec n, \Phi \right>_{\partial \Omega \setminus \partial \Omega_{S}} = 0. \label{eq:weak_modified_shallow_water_with_wetting_and_drying_continuity}
\end{align}\label{eq:weak_modified_shallow_water_with_wetting_and_drying}%
\end{subequations}
Here, $E$ denotes the interior mesh facets and
the superscripts $+$ and $-$ are used to distinguish between the two facet values for the discontinuous functions.
$\{\myvec u\}$ represents the downwind value of $\myvec u$, i.e.:
\begin{equation*}
\{ \myvec u \} \coloneqq
\begin{cases}
\myvec u \cdot \myvec n & \text{if } \myvec u \cdot \myvec n < 0, \\
   0       & \text{otherwise,}
\end{cases}
\end{equation*}
and $\llbracket \myvec u \rrbracket$ denotes
the jump of $\myvec u$ across the facet side:
\begin{equation*}
\llbracket \myvec u \rrbracket \coloneqq \myvec u^+ - \myvec u^-.
\end{equation*}
Note that the above formulation includes a simple upwinding scheme for the advection term, which is obtained by integrating the advection term by parts,
replacing the advected velocity at the inflow facets with the upwind velocity and then integrating by parts again.
The no-normal flow boundary condition has been weakly enforced by neglecting the surface integrals associated with the domain boundary $\partial \Omega_{S}$ in equation~\eqref{eq:weak_modified_shallow_water_with_wetting_and_drying_continuity}.
Similarly, the pressure term in the momentum equation~\eqref{eq:weak_modified_shallow_water_with_wetting_and_drying_momentum} is integrated twice by parts to weakly enforce the Dirichlet boundary condition on $\partial \Omega_{D}$.
As discussed in \citet{karna2011}, volume conservation is only satisfied if the integrals featuring the continuity equation~\eqref{eq:weak_modified_shallow_water_with_wetting_and_drying_continuity} are evaluated accurately.
Since $\tilde H$ is not a polynomial function, standard quadrature rules cannot evaluate these integrals exactly. Section~\ref{sec:wd_opt_verification_forward_model} investigates this issue and shows how the quadrature degree affects the volume conservation.
Another difficulty is to ensure that $\tilde H$ is positive everywhere also at the discrete level.
\citet{karna2011} uses piecewise linear elements for $\tilde H$ and exploits the fact that functions based on linear finite elements take their extrema at vertices.
Therefore a nodewise projection for $\tilde H$ ensures a domain-wide positive water level.
To circumvent this problem for the quadratic elements used here to represent water depth, $\tilde H$ itself is never stored as a discrete function, but is instead reevaluated for each quadrature point using equation~\eqref{eq:wd_opt_tilde_H_def}.


\subsection{Temporal discretisation}\label{sec:wd_opt_time_discretisation}
Following \cite{karna2011}, the weak equations \eqref{eq:weak_modified_shallow_water_with_wetting_and_drying} are discretised in time using the diagonally implicit Runge-Kutta scheme DIRK (2,3,2) \citep[\S 2.5]{ascher1997},
This is a second-order, L- and S-stable scheme, which allows for large time steps in the time integration.

The continuous time period is split into discrete levels with associated time steps $\Delta t$.
For each time level, DIRK schemes solve a sequence of stages, each of which requires solving a system of non-linear equations.
For brevity, we write the weak equations~\eqref{eq:weak_modified_shallow_water_with_wetting_and_drying} in the shortened form:
\begin{subequations}
\begin{align*}
& \left< \frac{\partial \myvec u}{\partial t}, \myvec \Psi\right>_\Omega = {S}_u(\eta, \myvec u), \\
& \left< \frac{\partial \tilde H}{\partial t}, \Phi \right>_\Omega = {S}_\eta(\eta, \myvec u).
\end{align*}
\end{subequations}
Let the superscript $n$ denote the time level and superscripts $i$ and $j$ denote DIRK stages.
The computation of time level $n$ involves the following steps:
\begin{itemize}
\item For each stage $i=1, \dots, s$ solve the following non-linear system for intermediate solutions $\myvec u^i$ and $\eta^i$:
\begin{subequations}
\begin{align*}
\left< \myvec u^i, \myvec \Psi\right>_\Omega & = \left< \myvec u^{n-1}, \myvec \Psi\right>_\Omega + \Delta t \sum_{j=1}^{i} a_{i,j} {S}_u(\eta^{j}, \myvec u^j),  \\
\left< \tilde H^i, \Phi \right>_\Omega & = \left< \tilde H^{n-1}, \Phi \right>_\Omega + \Delta t \sum_{j=1}^{i} a_{i,j} {S}_\eta(\eta^j, \myvec u^j).
\end{align*}
\end{subequations}
Each stage has an associated time level of $t^i = t^n + c_i \Delta t$ which is used to evaluate the forcing terms.
The coefficients $a_{i,j}$ and $c_i$ depend on the specific Runge-Kutta method and are defined below.
\item A final stage linearly combines the intermediate solutions to obtain the solution at the next time level $u^n$ and $\eta^n$:
\begin{subequations}
\begin{align*}
\left< \myvec u^n, \myvec \Psi\right>_\Omega & = \left< \myvec u^{n-1}, \myvec \Psi\right>_\Omega + \Delta t \sum_{j=1}^{s} b_j {S}_u(\eta^j, \myvec u^j),  \\
\left< \tilde H^n, \Phi \right>_\Omega & = \left< \tilde H^{n-1}, \Phi \right>_\Omega + \Delta t \sum_{j=1}^{s} b_j {S}_\eta(\eta^j, \myvec u^j).
\end{align*}
\end{subequations}
Again, the coefficients $b_j$ depend on the specific Runge-Kutta method used.
\end{itemize}
In general, the Runge-Kutta coefficients $a_{ij}, b_j$ and $c_i$ are defined compactly in the form of a Butcher tableau:
\vspace{1em}
\newline
\begin{tabular}{ l | c c c c }
  $c_1$ & $a_{1,1}$ \\
  $c_2$ & $a_{2,1}$ & $a_{2,2}$    \\
  \vdots & \vdots & \vdots  & $\ddots$ \\
  $c_s$ & $a_{s,1}$ & $a_{s,2}$  & \ldots & $a_{s,s}$ \\
\hline
  & $b_1$ & $b_2$ & \ldots & $b_s$ \\
\end{tabular}\\
\newline
The Butcher tableau for the DIRK (2,3,2) scheme used in this work is given by~\citep[\S 2.6]{ascher1997}:
\vspace{1em}
\newline
\begin{tabular}{ l | c c }
  $\gamma$ & $\gamma$ \\
  $1$ & $1-\gamma$ & $\gamma$ \\
\hline
  & $1-\gamma$ & $\gamma$ \\
\end{tabular}\\
\newline
with $\gamma \coloneqq \left(2-\sqrt{2}\right) / 2$.

\subsection{Verification}\label{sec:wd_opt_verification_forward_model}
The shallow water model with wetting and drying was implemented using the \fenics framework \citep{logg2011}.
The implementation was verified with the commonly used `Thacker' test case for which an analytical solution is known \citep{thacker1981}.

The Thacker test considers an undamped wave in a flat, bowl shaped basin where wetting and drying occurs on its sides.
The domain consists of a circular basin with a parabolic depth
\begin{equation*}
	h(x,y) \coloneqq h_c\left(1-\frac{x^2+y^2}{L^2}\right),
\end{equation*}
where  $L$ and $h_c$ are positive constants describing the basin's radius and depth at its centre, respectively.
\begin{figure}[p]
  \centering
  \includegraphics[width=0.6\textwidth]{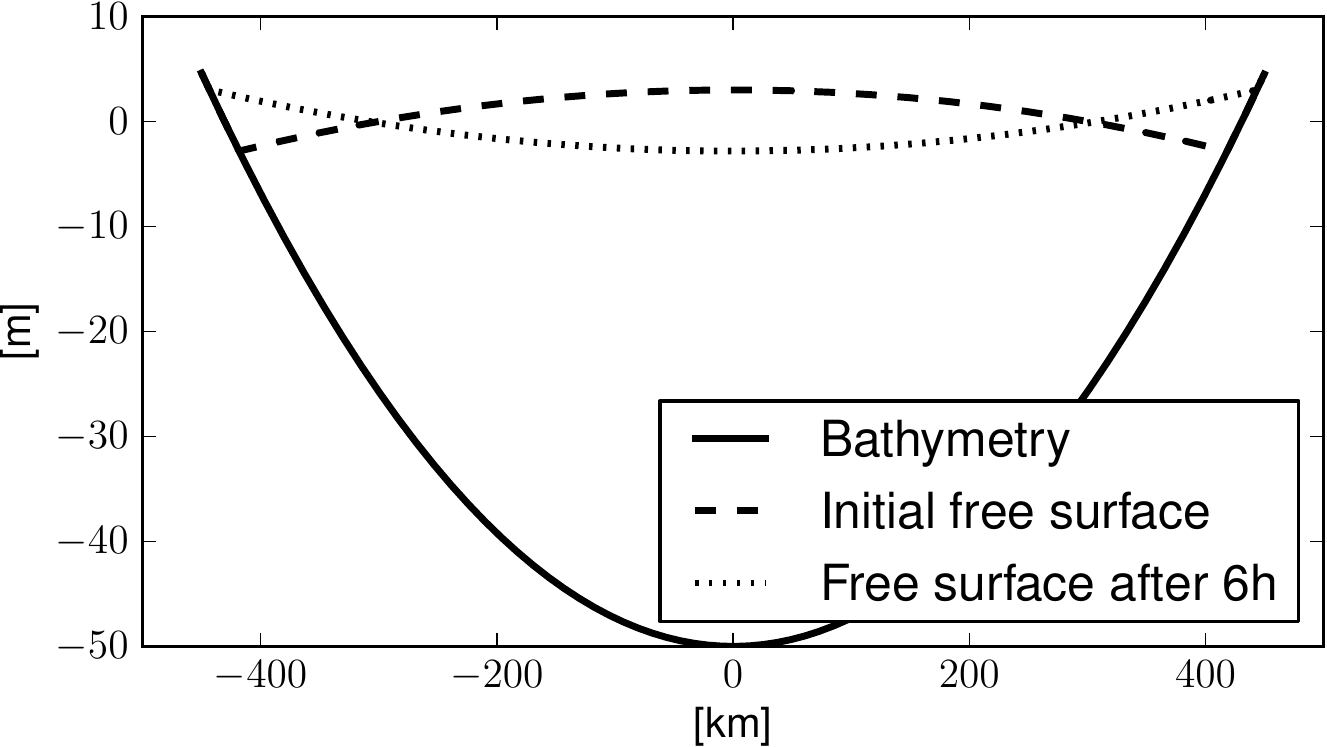}
  \caption{The setup of the Thacker problem. The free-surface oscillates with a $12$~h period, while wetting and drying occurs at the sides of the basin.}
  \label{fig:thacker_domain}
\end{figure}
The analytical solution satisfies the shallow water
equations with wetting and drying,
\eqref{eq:wd_opt_shallow_water_equations} and \eqref{eq:wetting_drying_extension},
without bottom friction, that is $\mu = 0$ and is:
\begin{align*}\label{eq:thacker_numerical_solution}
	\myvec u_{\textrm{exact}}(x,y, t) & \coloneqq \frac{\omega A \sin(\omega t)}{2(1-A\cos(\omega t))}
			    \begin{pmatrix}
			    x \\
			    y
			    \end{pmatrix}, \\
	\eta_{\textrm{exact}}(x,y, t) & \coloneqq h_c \left( \frac{\sqrt{1-A^2}}{1-A \cos \omega t}  - 1 - \frac{x^2 + y^2}{L^2} \left( \frac{1-A^2}{(1-A \cos \omega t)^2} -1 \right) \right),
\end{align*}
with
\begin{equation*}
	\omega^2\colonequals\frac{8gh_c}{L^2}, \qquad A\colonequals\frac{(h_c+\eta_c)^2-h_c^2}{(h_c+\eta_c)^2+h_c^2},
\end{equation*}
and $\eta_c$ is the maximum free-surface displacement at the basin's centre.
The parameters for the numerical tests were chosen to be consistent with \cite{balzano1998}:
$L=430.62$~km, $h_c=50$~m, $\eta_c=2$~m and a gravity magnitude of $g=9.81 \textrm{ m}/\textrm{s}^{2}$.
This results in a periodic free-surface oscillation with a $12$~h period, see figure~\ref{fig:thacker_domain}.

\begin{figure}[p]
\centering
        \begin{subfigure}[b]{0.4\textwidth}
                \centering
                \includegraphics[width=\textwidth]{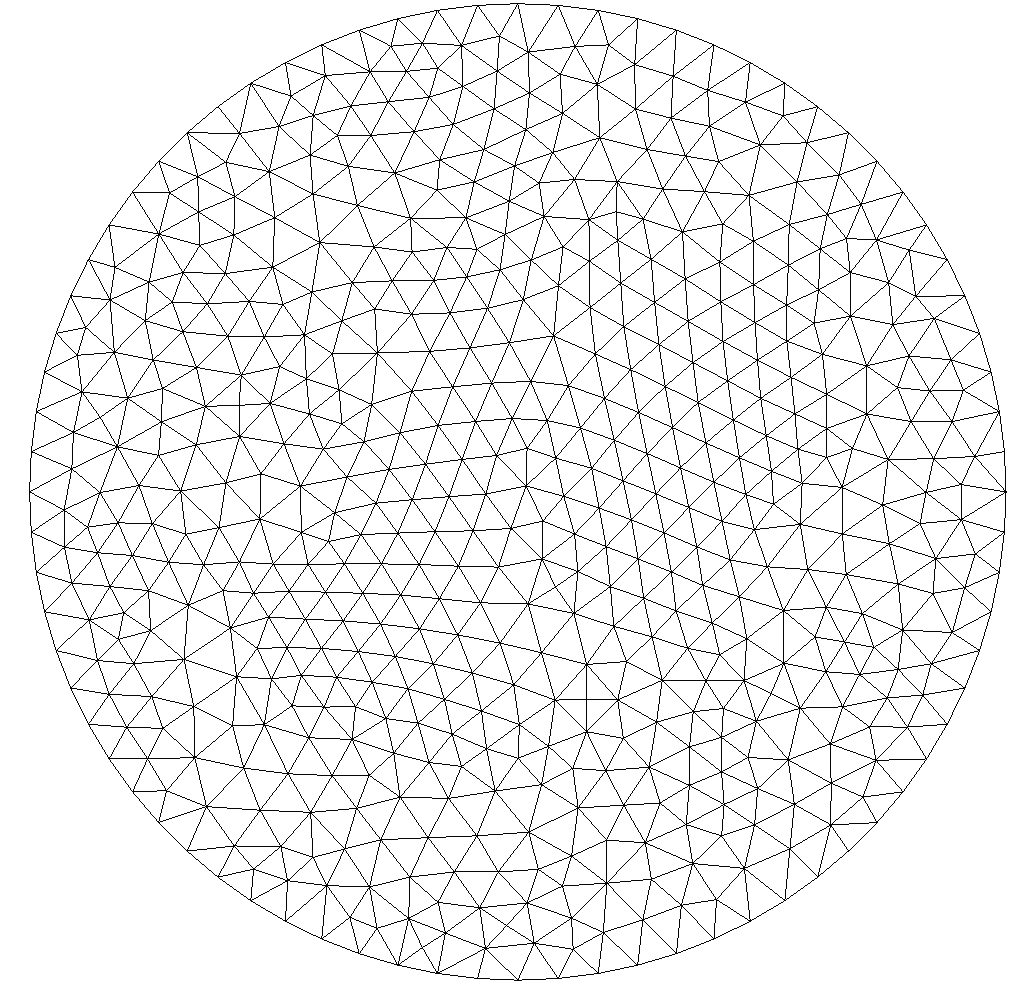}
                \caption{$40$~km mesh resolution}
                \label{fig:wd_optimal_control/pics/mesh40.png}
        \end{subfigure}
        ~
        \begin{subfigure}[b]{0.4\textwidth}
                \centering
                \includegraphics[width=\textwidth]{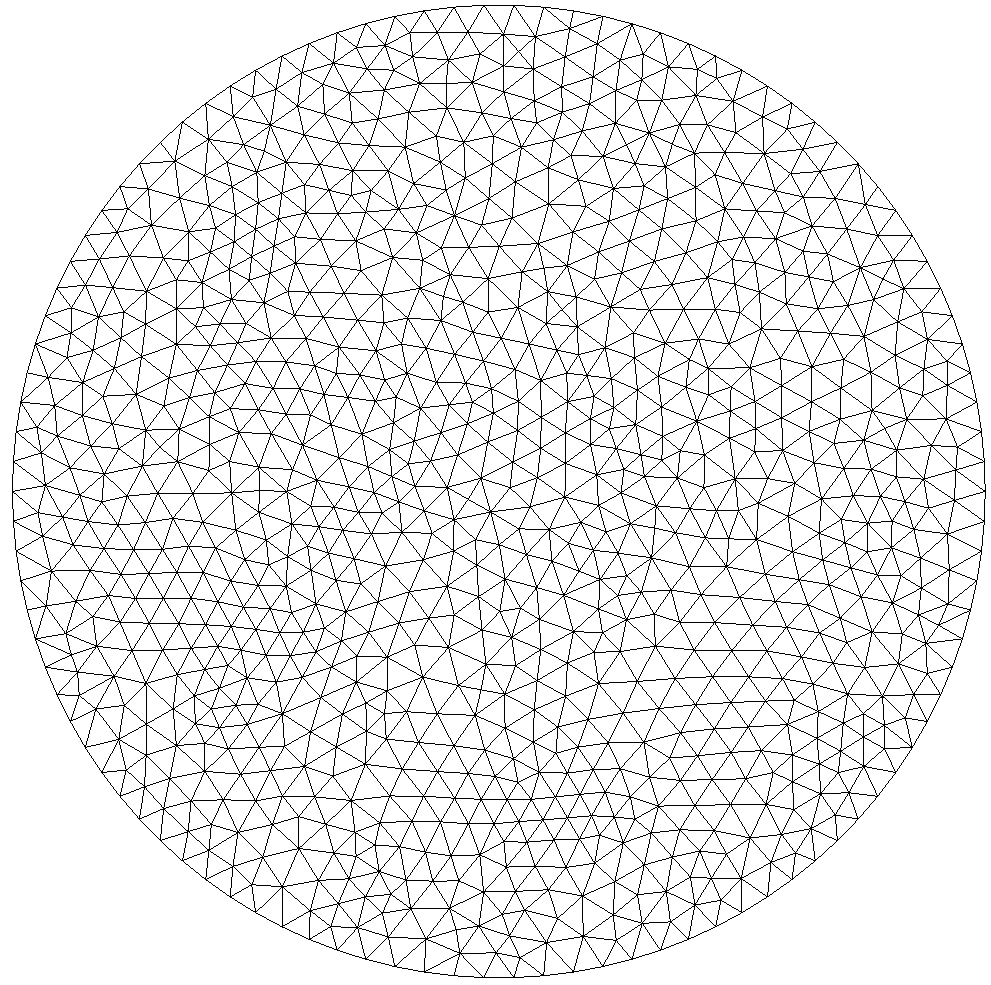}
                \caption{$30$~km mesh resolution}
                \label{fig:wd_optimal_control/pics/mesh30.png}
        \end{subfigure}
        ~
        \begin{subfigure}[b]{0.4\textwidth}
                \centering
                \includegraphics[width=\textwidth]{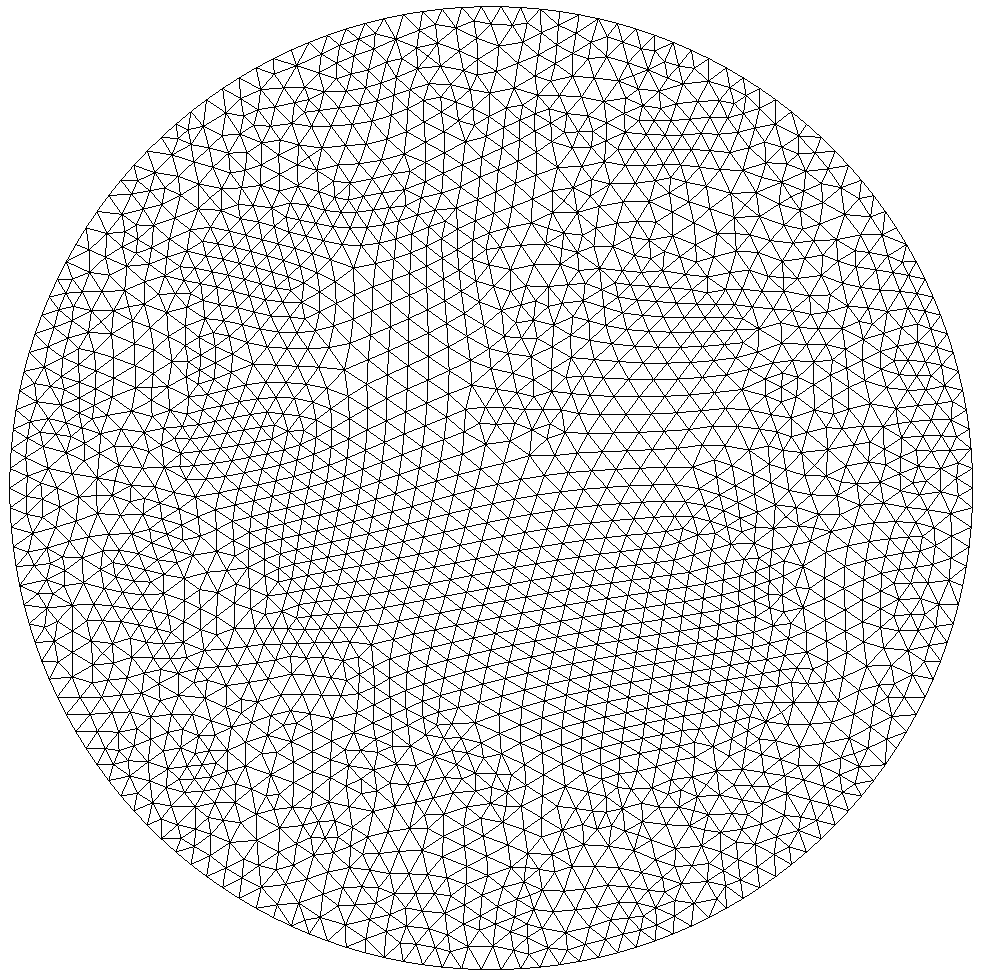}
                \caption{$20$~km mesh resolution}
                \label{fig:wd_optimal_control/pics/mesh20.png}
        \end{subfigure}
        ~
        \begin{subfigure}[b]{0.4\textwidth}
                \centering
                \includegraphics[width=\textwidth]{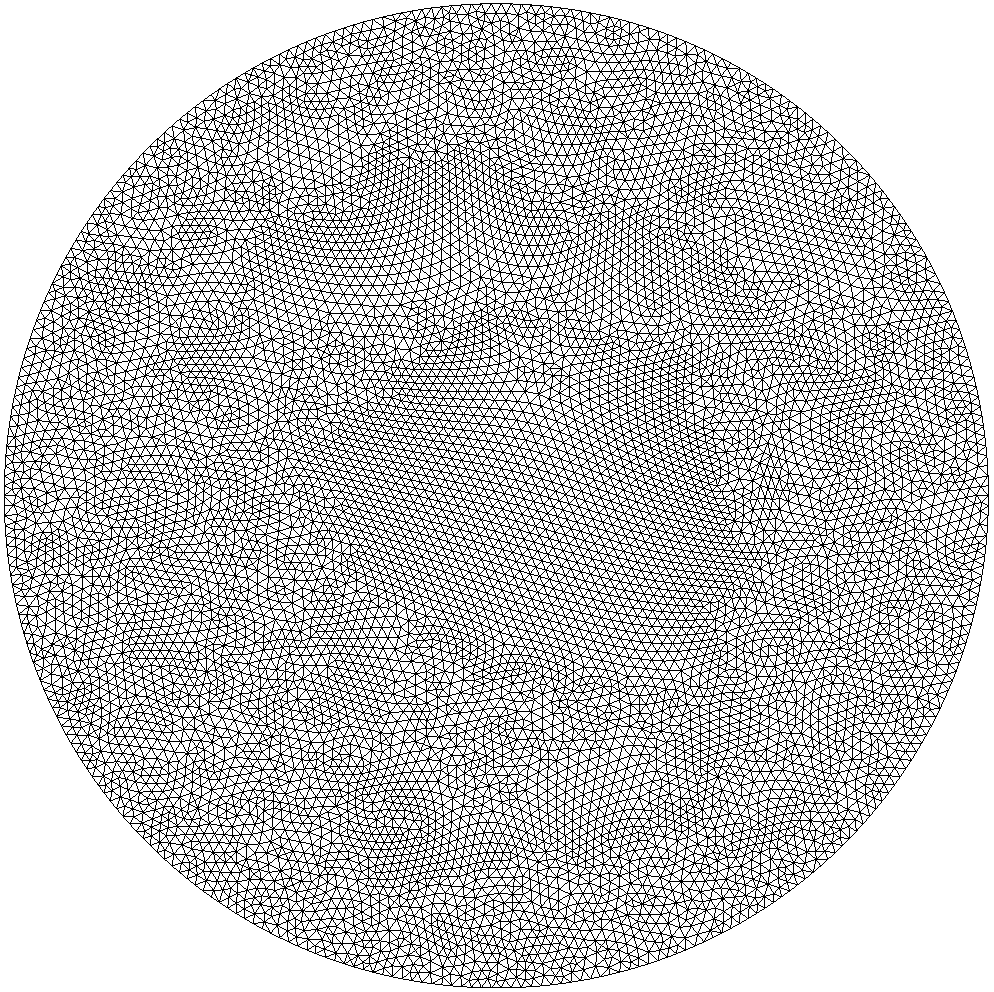}
                \caption{$10$~km mesh resolution}
                \label{fig:wd_optimal_control/pics/mesh10.png}
        \end{subfigure}
        \caption{The four meshes (generated with Gmsh~\citep{geuzaine2009}) used for the Thacker test case to determine the spatial convergence rate of the shallow water model with wetting and drying.}\label{fig:wd_optimal_control_thacker_meshes}
\end{figure}

The Thacker test case was numerically solved on four meshes with increasing resolution (figure~\ref{fig:wd_optimal_control_thacker_meshes}).
To ensure that the domain is sufficiently large to capture the wetting and drying process, the computational domain consists of a circle with radius $496.20$~km, in accordance to \cite{karna2011}.
The simulation was carried out for $24$~h with a  time step of $300$~s.
This time step is small enough to ensure that the spatial error dominates the temporal discretisation error:
performing the convergence analysis with a time step of $150$ s resulted in similar convergence results.
The smoothness constant $\alpha$ is estimated using equation~\eqref{eq:epsilon_definition_in_wd} and yields $\alpha \approx 2.4$~m for the finest mesh.
Subsequent numerical experiments showed that this value can further be reduced to $\alpha= 1.8$~m without compromising the stability of the simulation.
Hence, this reduced value was used for the finest mesh, and linearly increased with the mesh element sizes for the coarser meshes (that is $\alpha = 1.8, 3.6, 5.4, 7.2$ m for the $10, 20, 30, 40$ km element size meshes, respectively).

\begin{figure}[t]
\centering
  \includegraphics[width=0.5\textwidth]{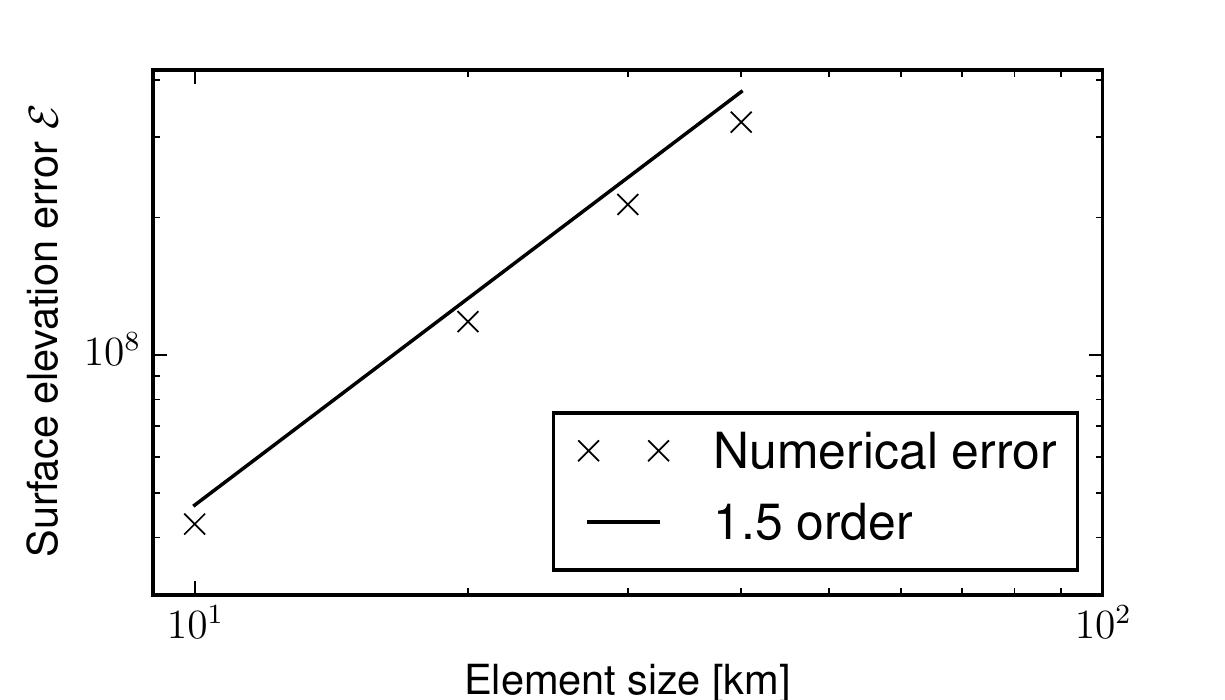}
  \caption{Spatial discretisation errors computed from the four meshes shown in figure~\ref{fig:wd_optimal_control_thacker_meshes}. The average rate of convergence is $1.46$;  \cite{karna2011} observed an order of $1.47$.}\label{fig:wd_optimal_rate_of_convergence_for_the_surface_elevation}
\end{figure}

The model implementation was verified by repeating the convergence test performed by~\citet{karna2011} and comparing the resulting order of convergence.
The error measure is defined as: 
\begin{equation*}
  \mathcal{E} \coloneqq \int_0^T \| \tilde \eta - \tilde \eta_{\textrm{exact}}\|_{L^2(\Omega)}\text{d}t,
\end{equation*}
where $\tilde \eta \coloneqq \tilde H - h$ and $\tilde \eta_{\textrm{exact}} \coloneqq \max(\eta_{\textrm{exact}}, -h)$ are the numerical and analytical solutions that take the bathymetry into account.
The numerical errors for the four meshes are plotted in figure~\ref{fig:wd_optimal_rate_of_convergence_for_the_surface_elevation}.
The average convergence rate is $1.46$, which is consistent with the observed order of convergence of $1.47$ in \citet{karna2011}.

\subsection{Volume conservation}
As discussed in section~\ref{sec:wd_opt_spatial_discretisation}, it has to be assessed whether, and to what extent, the finite quadrature rule affects volume conservation.
Furthermore, volume conservation might be affected by the tolerance setting of the Newton solver which is used to solve the non-linear problem at each timestep.

To study the conservation of volume, the Thacker test case introduced in the previous section was used.
It consists of a closed domain and therefore the fluid volume should remain constant throughout time.
Using the setup with the coarsest mesh from the previous section, the Thacker test case was solved for a combination of different quadrature degrees and Newton tolerances.
For each combination, the maximum relative error in the volume conservation was computed as
\begin{equation*}
  \mathcal{E}_V \coloneqq \max_{t\in(0, T)} \left|\frac{V(0)-V(t)}{V(0)}\right|,
\end{equation*}
where $V(t) \coloneqq \int_{\Omega} \tilde H(t) \dx$ is the total fluid volume at a time $t$.

The results of these tests are listed in table~\ref{table:wd_optimisation_thacker_mass_conservation_mesh_coarse}.
The volume conservation error is largely dominated by the tolerance of the Newton solver while the quadrature degree has only marginal influence.
The numerical simulations that follow use a quadrature degree of $20$ and a relative Newton solver tolerance of $10^{-9}$.

\begin{table}
\centering
  \begin{tabular}{c  c  | c }
Quadrature degree & Newton tolerance & Relative volume error $\mathcal E_V$ \\
      \hline
$4$ & $10^{-6}$ & $1.3 \times 10^{-8}$ \\
$4$ & $10^{-9}$ & $5.7 \times 10^{-12}$ \\
$4$ & $10^{-12}$ & $4.5 \times 10^{-15}$ \\
      \hline
$20$ & $10^{-6}$ & $1.3 \times 10^{-8}$ \\
$20$ & $10^{-9}$ & $5.7 \times 10^{-12}$ \\
$20$ & $10^{-12}$ & $2.8 \times 10^{-15}$ \\
      \hline
$121$ & $10^{-6}$ & $1.3 \times 10^{-8}$ \\
$121$ & $10^{-9}$ & $5.7 \times 10^{-12}$ \\
$121$ & $10^{-12}$ & $2.7 \times 10^{-15}$ \\
  \end{tabular}
\caption{The maximum relative volume conservation error over $24$~h on the coarsest mesh (figure \ref{fig:wd_optimal_control/pics/mesh40.png}) for different quadrature degrees and relative Newton solver tolerances.}
\label{table:wd_optimisation_thacker_mass_conservation_mesh_coarse}
\end{table}

\subsection{Validation}
The validation of the forward model is outside the scope of this work.
However, it should be noted that the wetting and drying scheme employed here has previously been applied to the Scheldt estuary and the North Sea, and validated against tidal stations with good results in \citet{karna2011} and \citet{gourgue2011}.

\section{The data assimilation problem}
\subsection{Formulation as an optimisation problem}
This section formulates the problem of reconstructing the profile of an incoming wave from inundation observations as an optimisation problem constrained by the shallow water equations.
This will allow us to apply techniques such as gradient-based optimisation methods and the adjoint model to efficiently solve the data assimilation problem.

The goal quantity that we aim to minimise measures the misfit between
an observed and the simulated wet/dry interface at all time levels.
For that, we map the simulated water height $\eta$ to an
indicator function which approaches $1$ in dry and $0$ in wet areas. By noting
that $\eta \ge h$ in wet and $\eta < h$ in dry areas (see
figure~\ref{fig:wd_optimisation/pics/wd_explained_fixed}), this indicator
function is defined as $\mathcal{H}(\eta - h)$  where
$\mathcal{H}$ is a smooth approximation of the Heaviside step function:
\begin{equation}
  \mathcal{H}(x) \coloneqq \frac{1}{2}\left( \frac{x}{\sqrt{x^2 + \alpha^2}} + 1 \right)\approx
\begin{cases}
0 & \mbox{if } x < 0, \\
1 & \mbox{otherwise},
\end{cases} \label{eq:wd_opt_smooth_heaviside_approx}
\end{equation}
where $\alpha$ controls the smoothness of the approximation.
A plot of this approximation in given in figure~\ref{fig:wd_optimisation/pics/plot_heaviside}.
Given some observations $d$ of a wet/dry interface (that is, $d$ is an time-varying function that approaches $1$ at dry and $0$ at wet points), we define the goal quantity as
\begin{equation}
    J(\eta, \eta_D) \coloneqq \frac{1}{2} \int_0^T \int_\Omega \left|\mathcal{H}(\eta - h) - d\right|^2 \dx\dt + \frac{\beta}{2} \int_0^T \int_{\partial \Omega_D} \left|\frac{\partial \eta_D}{\partial t}\right|^2 \dx \dt.
\label{eq:opt_wd_functional} 
\end{equation}
The first term quantifies the discrepancy between simulated and observed wet/dry interfaces, while the second term is a regularisation term that enforces temporal smoothness in the boundary displacement: a larger $\beta$ value results in smoother
boundary displacement in the reconstructed profile.

The optimisation parameters are the Dirichlet boundary values $\eta_D$ at each
time level in the shallow water
equations~\eqref{eq:modified_shallow_water_with_wetting_and_drying}.  For
simplicity, in our computations it is assumed that the boundary values only vary in time and are
constant in space. For spatially varying boundary conditions, the functional \ref{eq:opt_wd_functional} would need to be extended by an additional spatial regularisation term. Note that the computation of the Runge-Kutta stages requires the Dirichlet boundary values at intermediate time levels, see
section~\ref{sec:wd_opt_time_discretisation}.  These values are obtained by
linearly interpolating the boundary values from the two neighbouring time levels.

We can now state the data assimilation problem as an optimisation problem with the shallow water equations as a constraint:
\begin{equation}
\begin{aligned}
\min_{\eta_D,u,\eta}\ J(\eta, \eta_D) \quad & \mbox{subject to} \\
  \frac{\partial \myvec u}{\partial t} + (\myvec u \cdot \nabla) \myvec u + g \nabla \eta &= -\frac{c_f(\tilde H)}{\tilde H} \|\myvec u\|\myvec u && \mbox{on } \partial \Omega \times (0, T), \\
\frac{\partial \eta}{\partial t} + \frac{\partial \tilde h}{\partial t} + \nabla \cdot  (\tilde H \myvec u) &= 0 && \mbox{on } \partial \Omega \times (0, T),  \\
\myvec u \cdot \myvec n & = 0 && \mbox{on } \partial \Omega_{S} \times (0,T),\\
\eta & = \eta_{D} && \mbox{on }  \partial \Omega_{D} \times (0, T), \\
\myvec u & = \myvec u_0, \quad \eta = \eta_0 && \mbox{at } \Omega \times \{0\}.
\end{aligned}\label{eq:modified_shallow_water_with_wetting_and_drying_optimisation_problem}%
\end{equation}

\subsection{Adjoint model implementation}
In order to efficiently solve the optimisation
problem~\eqref{eq:modified_shallow_water_with_wetting_and_drying_optimisation_problem},
the total derivative of the goal quantity with respect to the
optimisation parameters, $\textrm{d}J / \text{d}\eta_D$, is required.
This quantity is here computed by solving the adjoint equations.

For the mathematical derivation of the adjoint shallow water equations we refer
the reader to \citet[\S 5.4.4 and appendix C]{funke2012}. The adjoint model was
automatically generated using the \fenics extension \da \citep{farrell2012}.
\da derives the adjoint model, and the derivative computation, directly from
the discretised shallow water equations. This has the advantage that the
derivative is also the derivative of the discrete shallow water model, rather
than just another approximation of the non-discrete derivative.  Without
this discrete consistency, the derivative might a bad descent direction for the
optimisation, and one would need to use a more robust optimisation methods.

The adjoint implementation was verified with the Taylor remainder convergence test \citep[\S 2.5.6]{funke2012}. Its
application to a simple example yielded the expected first-order convergence without gradient information,
and second-order convergence with gradient information.
Furthermore, the Taylor remainder convergence test was successfully applied to
the first ten optimisation iterations of all numerical examples in this paper.
This gives high confidence that the adjoint system and the gradient computation
are correctly implemented.

\section{Numerical examples}
This section performs numerical experiments on two data assimilation problems with inundation observations.
In both cases, the resulting optimisation problems \eqref{eq:modified_shallow_water_with_wetting_and_drying_optimisation_problem} were solved with the limited memory BFGS method with bound support (L-BFGS-B) from SciPy~\citep{byrd1995, scipy}.
The L-BFGS-B method belongs to the class of quasi-Newton algorithms that use an approximation of the Hessian matrix based on a limited number of functional gradients (here 10).

In order to be able to investigate the effectiveness and robustness of the wave reconstruction, we apply the 
method to synthetically generated observations of the wet/dry interface. 
The synthetic observations were obtained by first choosing a Dirichlet boundary condition $\eta_D^{\textrm{exact}}$, then solving the shallow water model, and recording the wet/dry interface at each timestep as $\mathcal H(\eta-h)$.
Using these records as the observations $d$ in the goal
quantity~\eqref{eq:opt_wd_functional} guarantees that the chosen Dirichlet
boundary condition $\eta_D^{\textrm{exact}}$ is a solution to the optimisation
problem~\eqref{eq:modified_shallow_water_with_wetting_and_drying_optimisation_problem}.

\subsection{Wave profile reconstruction on a sloping beach}
The first data assimilation problem consists of a long, thin sloping beach with an incoming wave on the deep side.
The goal is to reconstruct the wave profile based on observations of the wet/dry interface.

\begin{figure}[bt]
\centering
  \begin{subfigure}[b]{0.48\textwidth}
      \centering
      \includegraphics[width=0.9\textwidth]{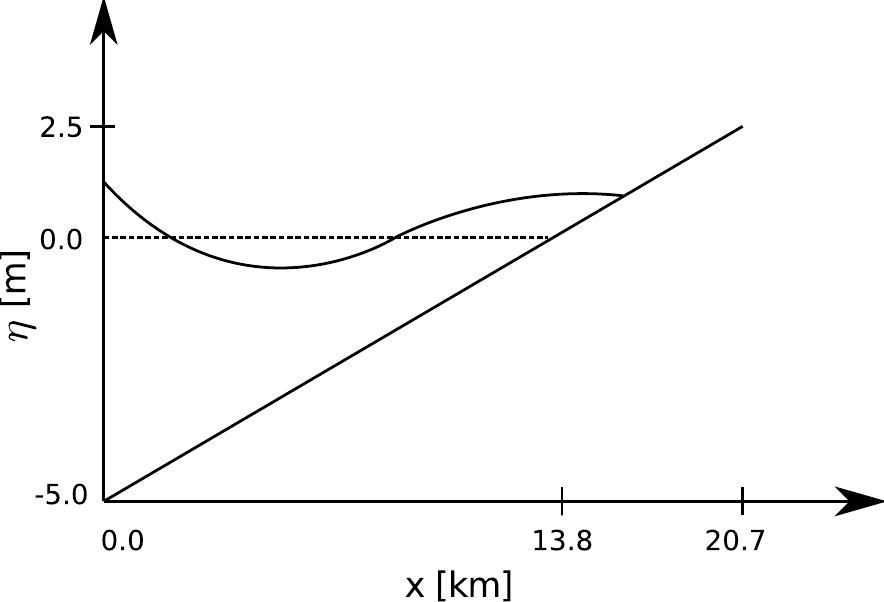}
      \caption{Domain}\label{fig:wd_optimal_rate_optimal_control_domain}
  \end{subfigure}
  \begin{subfigure}[b]{0.48\textwidth}
      \centering
      \includegraphics[width=\textwidth]{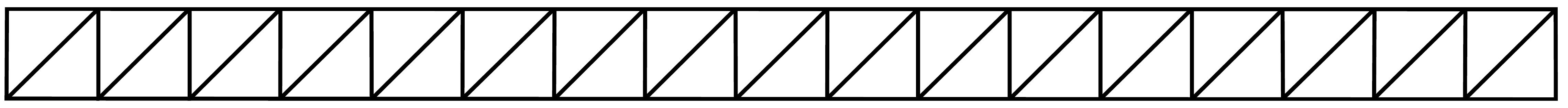}
      \vspace{1cm}
      \caption{Mesh}\label{fig:wd_optimal_rate_optimal_balzano_mesh}
  \end{subfigure}
  \caption{A vertical slice of the setup for the wave profile reconstruction on a sloping beach.}\label{fig:wd_optimal_rate_optimal_balzano_setup}
\end{figure}

The computational domain is an adaption of a wetting and drying test case considered by \citet{balzano1998}.
It consists of a linearly increasing slope of $1.2$~km width and $20.7$~km length.
The left end of the slope is $5$~m below, and the right end $2.5$~m above the reference water level, see figure~\ref{fig:wd_optimal_rate_optimal_control_domain}.
The Dirichlet boundary condition $\eta_D$ controls the water level on the left domain boundary, and a no-normal flow is enforced on the other boundaries.

The remaining parameters are $g = 9.81 \textrm{ m}/\textrm{s}^2$ for the gravity constant and $\mu = 0.025$ s$/$m$^{1/3}$ for the Manning drag coefficient.
The domain is uniformly discretised with triangular elements of $1.2$~km size, see figure~\ref{fig:wd_optimal_rate_optimal_balzano_mesh}.
For this mesh, the guideline equation for the smoothness parameter~\eqref{eq:epsilon_definition_in_wd} suggests a value of $\alpha = 0.43$~m.
The time step is set to $\Delta t = 600$~s and the final time is $T=24$~h.

The model inputs to be reconstructed by the data assimilation are chosen to be the free-surface displacement values on left boundary for each time level, except during the final $2$ h.
The final 2 h cannot be reconstructed because the boundary values have no influence on the wet/dry interface due to the finite wave speed.

\subsubsection{Sinusoidal wave profile without noisy observations}\label{sec:sin_no_noise}
\begin{figure}[p]
\centering
        \begin{subfigure}[b]{0.48\textwidth}
                \centering
		\includegraphics[width=1\textwidth]{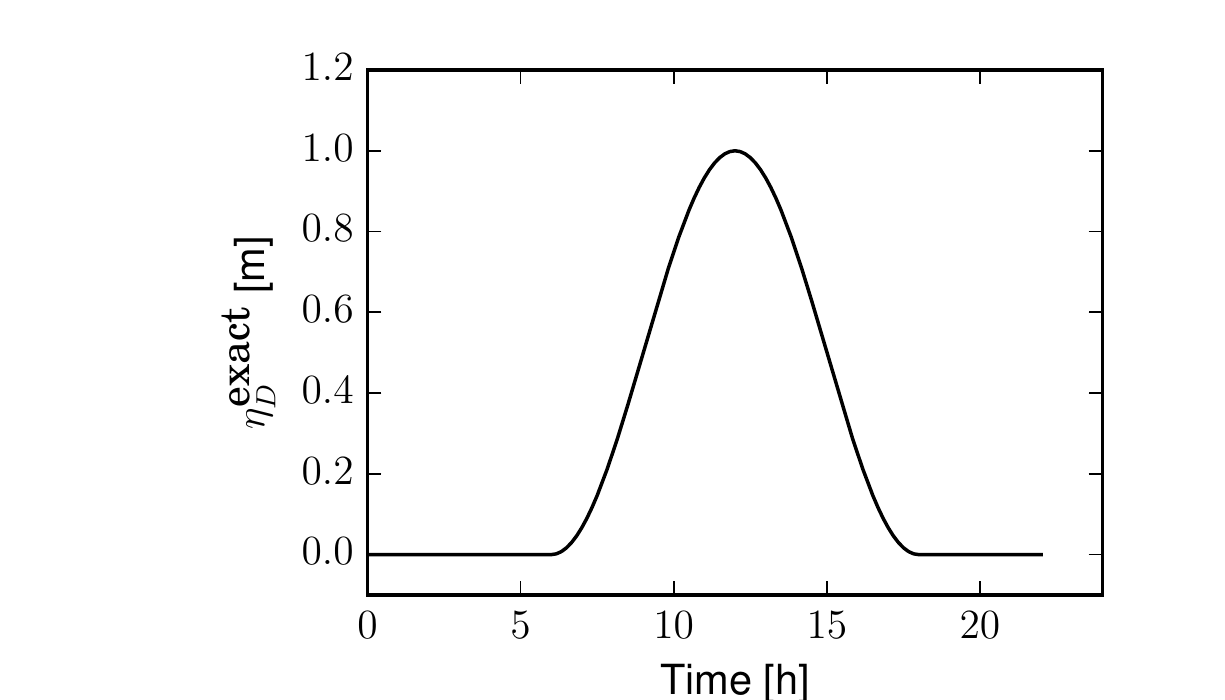}
		\caption{Reference wave profile\\~}\label{fig:wetting_drying_balzano_0.43_controls_optimal}
        \end{subfigure}
        \begin{subfigure}[b]{0.48\textwidth}
                \centering
		\includegraphics[width=1\textwidth]{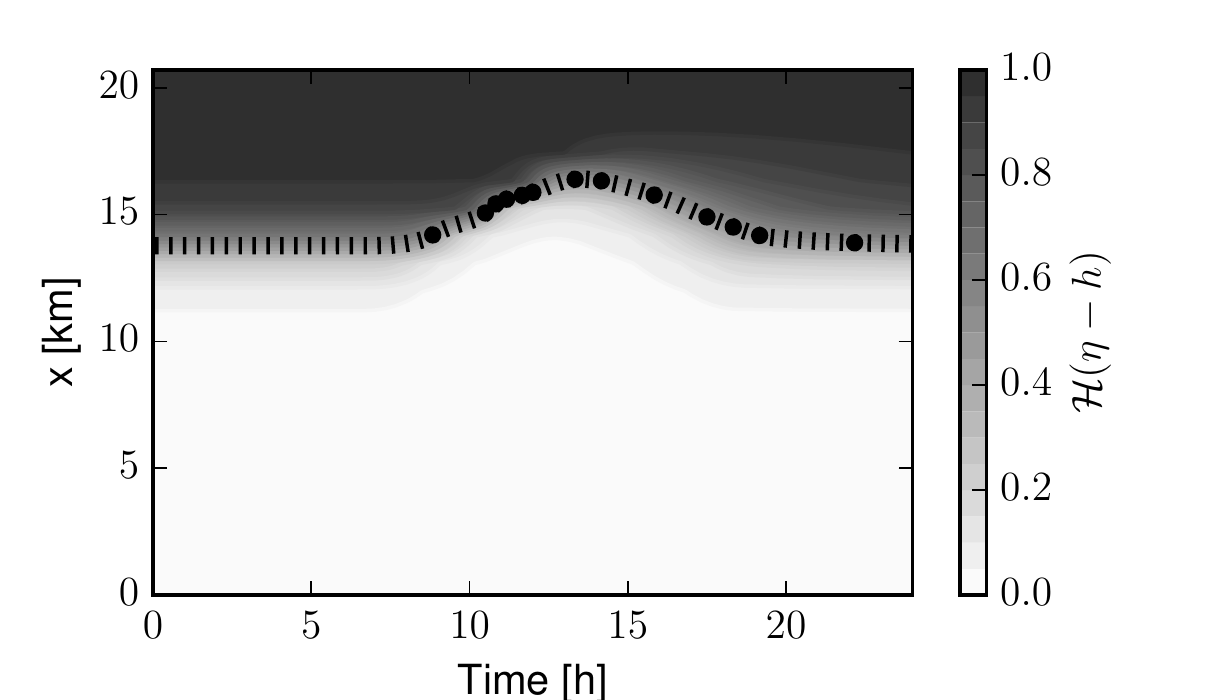}
		\caption{Observations and exact wet/dry interface (dashed line)}\label{fig:wetting_drying_balzano_0.43_plot_wd_interface}
        \end{subfigure}\\
        \begin{subfigure}[b]{0.48\textwidth}
                \centering
		\includegraphics[width=1\textwidth]{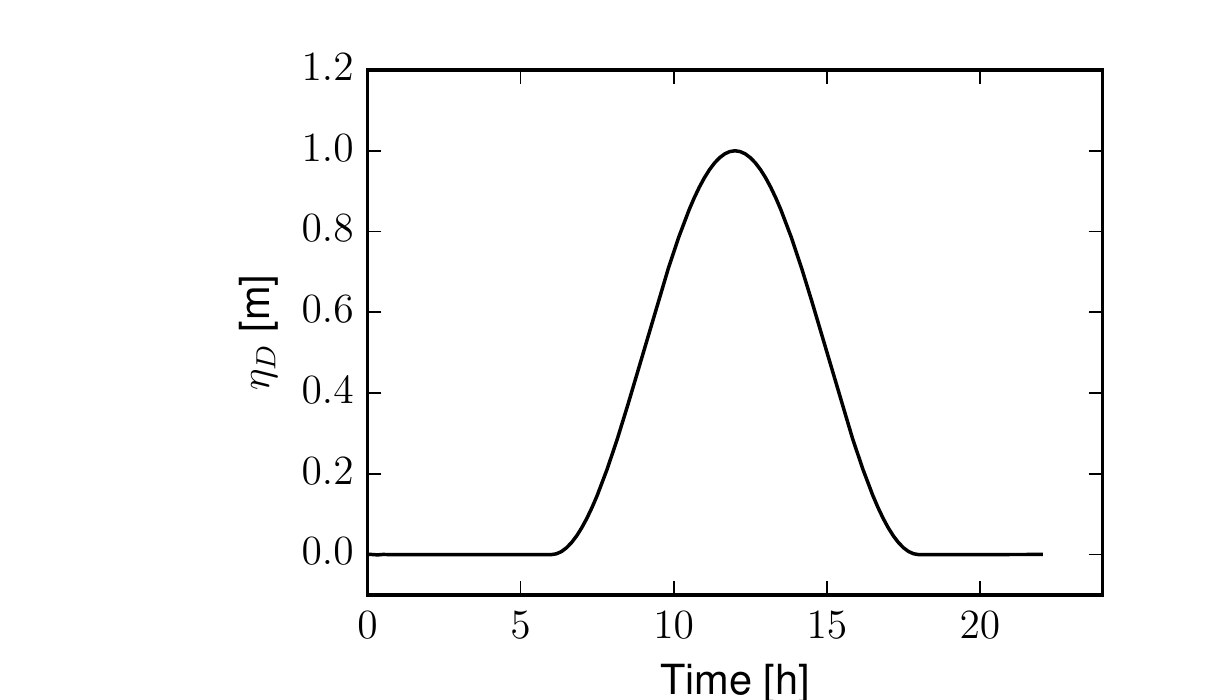}
		\caption{Reconstructed wave profile}\label{fig:wetting_drying_balzano_0.43_controls_final}
        \end{subfigure}
        \begin{subfigure}[b]{0.48\textwidth}
                \centering
		\includegraphics[width=1\textwidth]{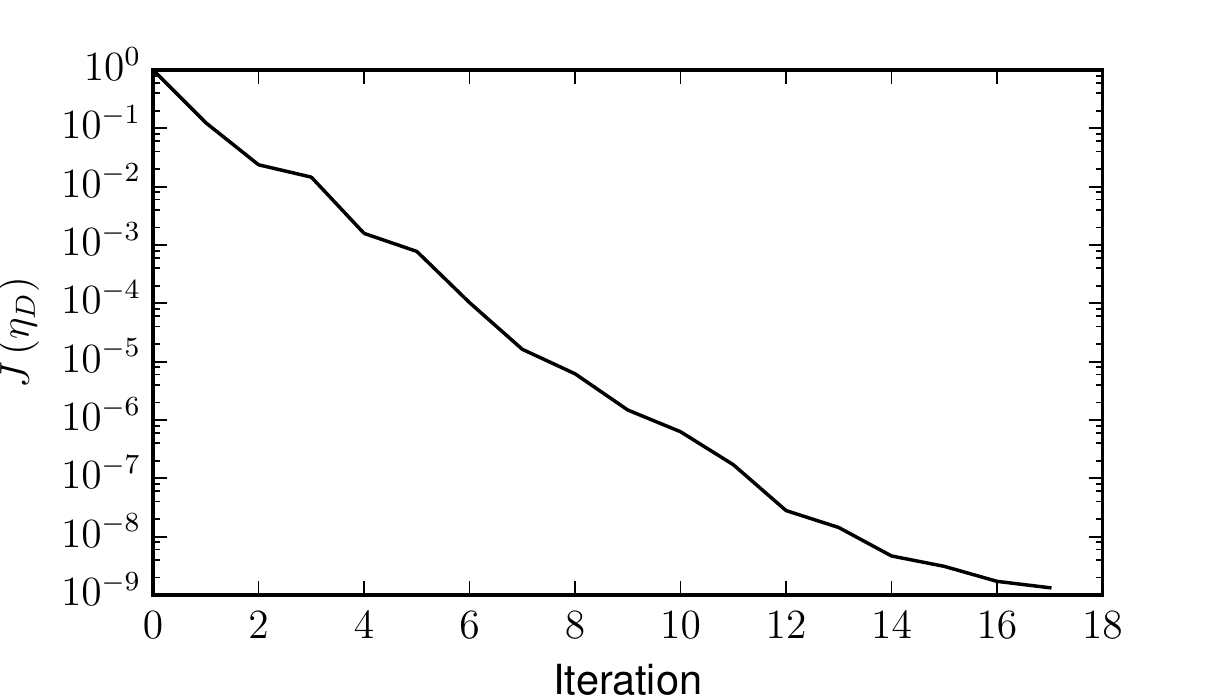}
		\caption{Optimisation convergence}\label{fig:wetting_drying_balzano_0.43_iter_plot}
        \end{subfigure}
        \caption{Results of the noiseless wave profile reconstruction on a sloping beach with a sinusoidal incoming wave profile and smoothing value $\alpha = 0.43$~m.
                 The final $2$ h of the wave profile are excluded from the reconstruction.}
\label{fig:wd_balzano1_results_alpha_0.43}
\end{figure}

%
In a first experiment, the reference Dirichlet boundary $\eta_D^{\textrm{exact}}$ consists of
one sinusoidal wave with $p\coloneqq12$~h period and $1$~m amplitude (figure~\ref{fig:wetting_drying_balzano_0.43_controls_optimal}):
\begin{equation}
\eta_{D}^{\textrm{exact}}(t) \coloneqq
\begin{cases}
  0 & \textrm{if } t < 6 \textrm{ h}, \\
\frac{1}{2}\left(-\cos\left(\frac{2\pi (t - 6 \textrm{ h})}{p}\right) + 1\right) & \textrm{if } 6 \textrm{ h} \le t \le 18 \textrm{ h}, \\
0 & t > 18 \textrm{ h.}
\end{cases}
\label{wd:optimal_control_choice_simple_sinusoidal_wave}
\end{equation}
The observations $d$ in the goal quantity~\eqref{eq:opt_wd_functional} are generated by solving the shallow water model with these boundary values for $24$~h and recording the wet/dry interface $\mathcal H(\eta - h)$ at every time level.
The resulting observations with $\alpha = 0.43$~m are plotted in figure~\ref{fig:wetting_drying_balzano_0.43_plot_wd_interface}.

To verify that the reference Dirichlet boundary condition $\eta_D^{\textrm{exact}}$ combined with these observations is indeed a solution to the optimisation problem~\eqref{eq:modified_shallow_water_with_wetting_and_drying_optimisation_problem}, the
optimisation algorithm was executed with regularisation coefficient $\beta=0$ and $\eta_D = \eta_D^{\textrm{exact}}$ as an initial guess.
As expected, the algorithm terminated after the first iteration, reporting that the first-order optimality conditions hold (i.e. the gradient of the goal quantity is zero).

Next it was tested if the reference Dirichlet boundary condition can be reconstructed without prior information.
For that, the optimisation problem~\eqref{eq:modified_shallow_water_with_wetting_and_drying_optimisation_problem} was solved with $\beta=0$ and an initial guess of $\eta_D = 0$.
The optimisation algorithm was terminated once the relative change of the functional value in one optimisation iteration dropped below $10^{-9}$.
With that setup, the optimisation finished after 17 iterations, see figure \ref{fig:wetting_drying_balzano_0.43_iter_plot}.
Figure \ref{fig:wetting_drying_balzano_0.43_controls_final} shows that
the reference Dirichlet condition was accurately reconstructed. The maximum discrepancy
of the incoming wave profile is $0.1$ cm, which
corresponds to a relative reconstruction error of $0.1$ \%.

\begin{figure}[p]
\centering
        \begin{subfigure}[b]{0.48\textwidth}
                \centering
		\includegraphics[width=1\textwidth]{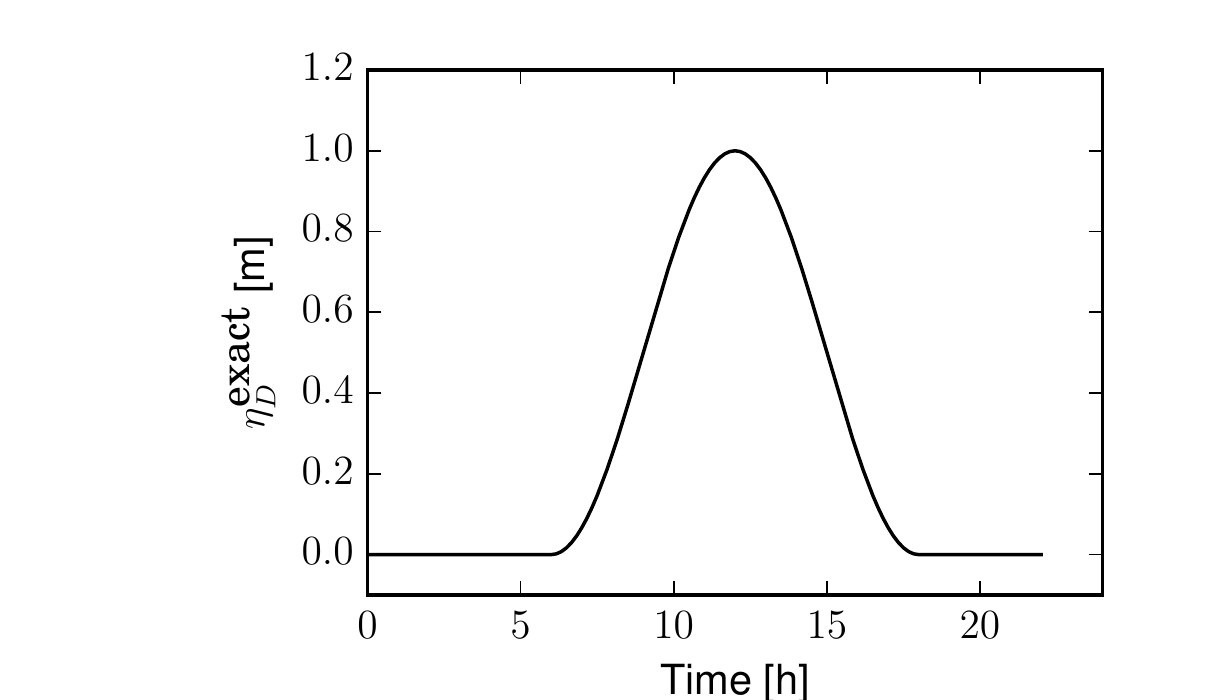}
		\caption{Reference wave profile\\~}\label{fig:wetting_drying_balzano_1.8_controls_optimal}
        \end{subfigure}
        \begin{subfigure}[b]{0.48\textwidth}
                \centering
		\includegraphics[width=1\textwidth]{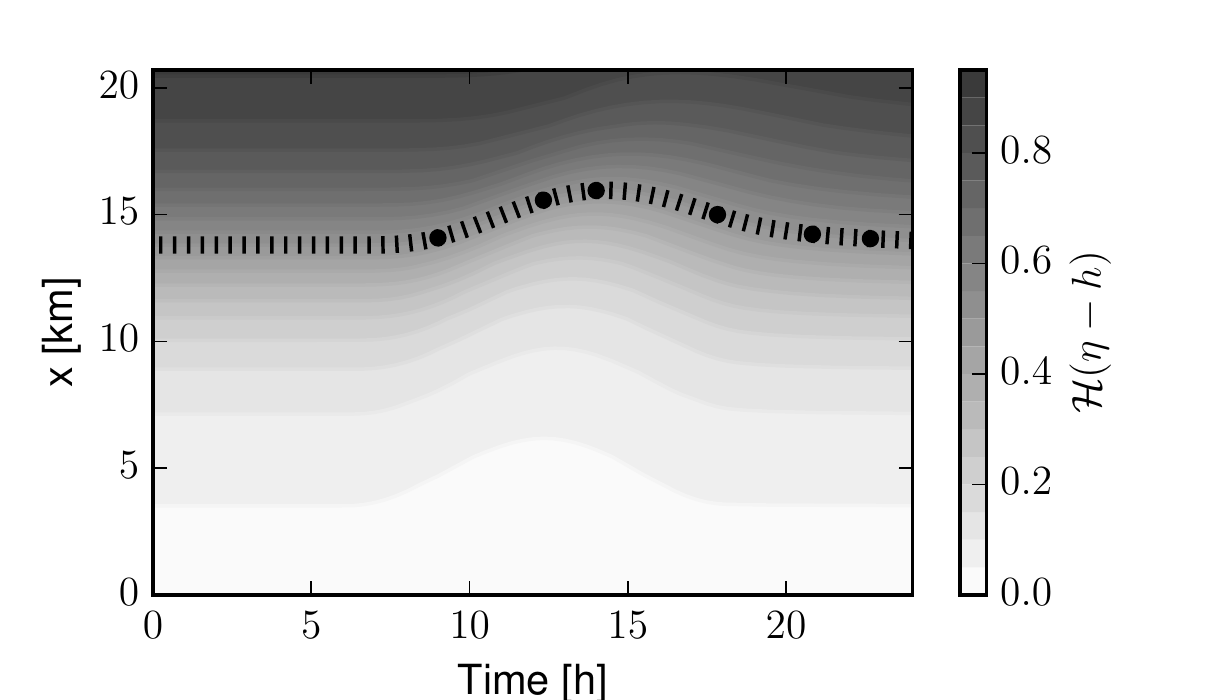}
            \caption{Observations and exact wet/dry interface (dashed line)}\label{fig:wetting_drying_balzano_1.8_plot_wd_interface}
        \end{subfigure}
	\\
        \begin{subfigure}[b]{0.48\textwidth}
                \centering
		\includegraphics[width=1\textwidth]{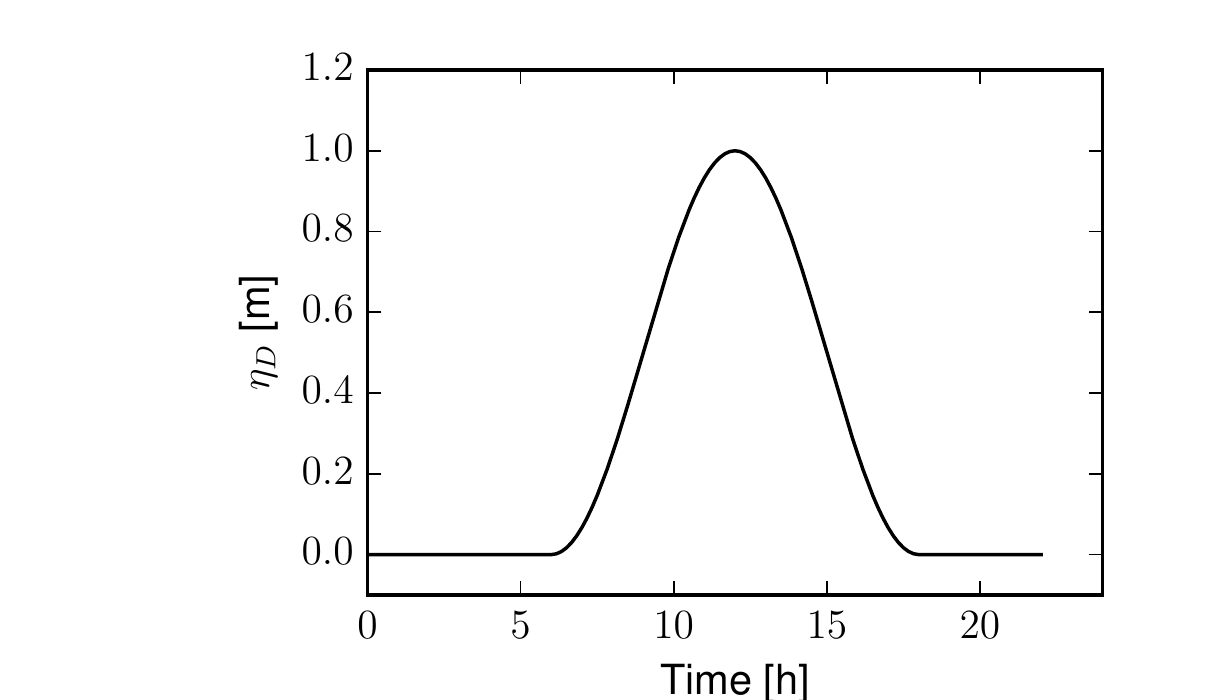}
		\caption{Reconstructed wave profile}\label{fig:wetting_drying_balzano_1.8_controls_final_reconstructed}
        \end{subfigure}
        \begin{subfigure}[b]{0.48\textwidth}
                \centering
		\includegraphics[width=1\textwidth]{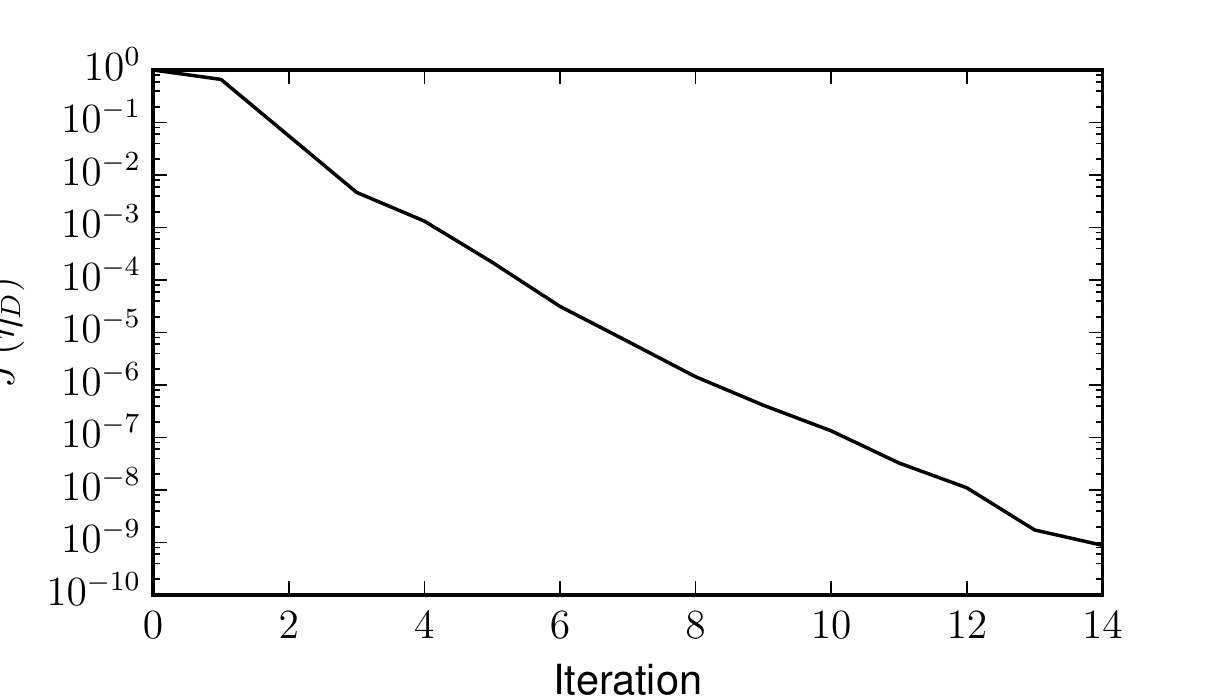}
		\caption{Optimisation convergence}\label{fig:wetting_drying_balzano_1.8_iter_plot}
        \end{subfigure}
        \caption{Results of the noiseless wave profile reconstruction on a sloping beach with a sinusoidal incoming wave profile and smoothing value $\alpha = 1.8$~m.
                 The final $2$ h of the wave profile are excluded from the reconstruction.}\label{fig:wd_balzano1_results_alpha_1.8}
\end{figure}

To test the impact of the smoothing parameter we increased $\alpha$ from
$0.43$~m to $1.8$~m.  The results are shown in
figure~\ref{fig:wd_balzano1_results_alpha_1.8}. Compared to the previous
experiment, the smoother gradient at the wet/dry interface can clearly be seen.
With this setup, the optimisation algorithm terminated after $14$ iterations and
reconstructed the incoming wave profile up to a maximum error of less than
$0.03$~cm.  Overall, the reconstruction seems to work well for different $\alpha$
values, however, a large smoothing constant can cause unphysical results as
described in \citet{karna2011}.

\subsubsection{Sinusoidal wave profile with noisy observations}

\begin{figure}[p]
    \centering
        \begin{minipage}[t]{.48\textwidth}
                \centering
            a) $\sigma=0.1$
        \end{minipage}
        \begin{minipage}[t]{.48\textwidth}
                \centering
            b) $\sigma=0.5$
        \end{minipage}
        \\
        \begin{minipage}[t]{.49\textwidth}
            \flushright
            \begin{subfigure}[b]{0.85\textwidth}
            \flushright
    		\includegraphics[width=\textwidth]{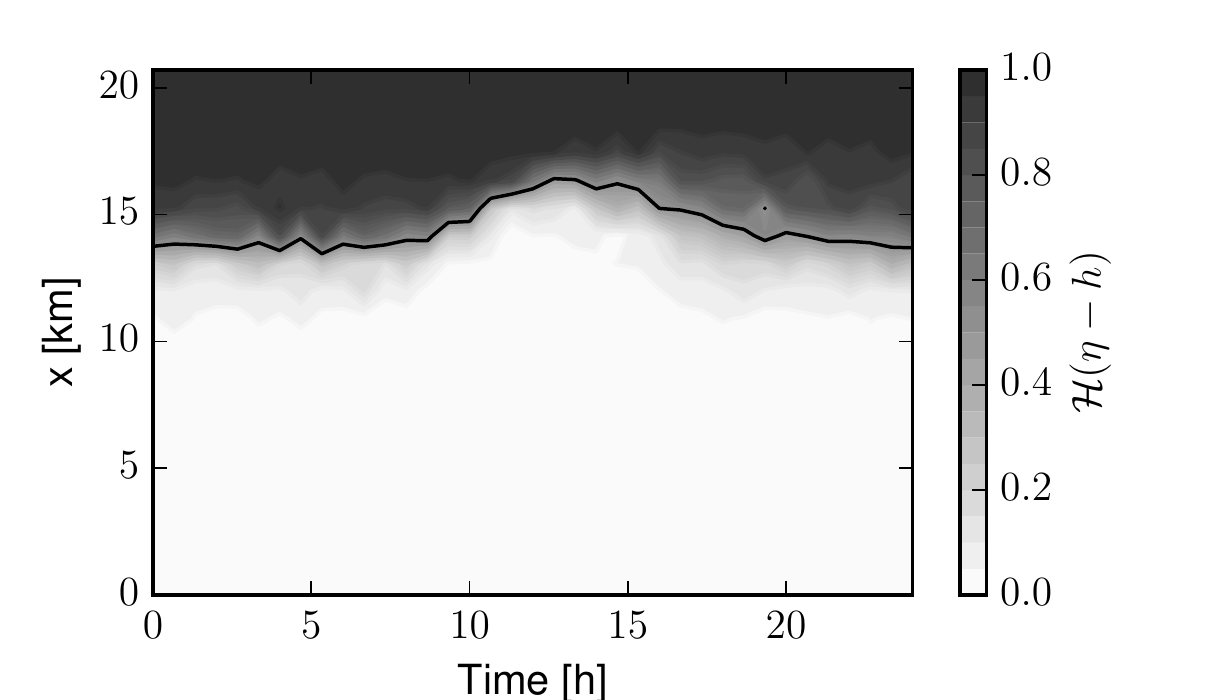}
    		\caption*{Observations and wet/dry interface}
            \label{fig:wetting_drying_balzano_0.43_noise_results_dg_noise_0.1_plot_wd_interface}
            \end{subfigure}
        \end{minipage}
        \begin{minipage}[t]{.49\textwidth}
            \flushright
            \begin{subfigure}[b]{0.85\textwidth}
            \flushright
    		\includegraphics[width=\textwidth]{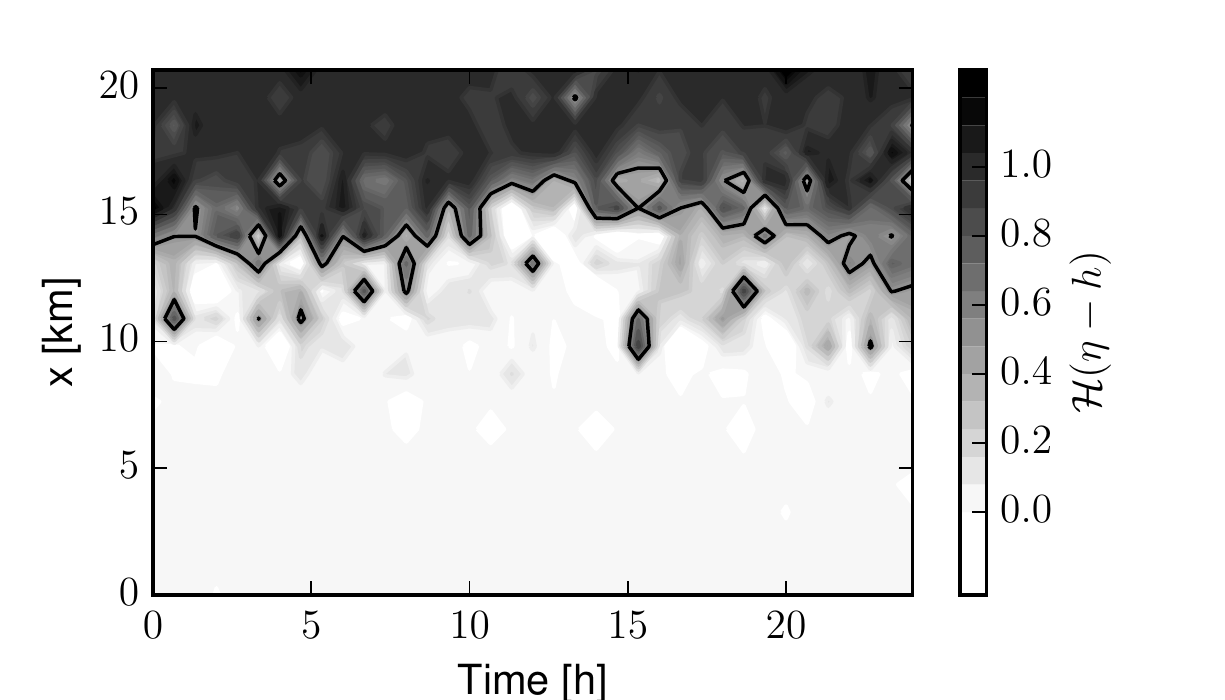}
    		\caption*{Observations and wet/dry interface}\label{fig:wetting_drying_balzano_0.43_noise_results_dg_noise_0.5_plot_wd_interface}
            \end{subfigure}
        \end{minipage}
        \\
        \begin{subfigure}[b]{0.48\textwidth}
                \centering
		\includegraphics[width=1\textwidth]{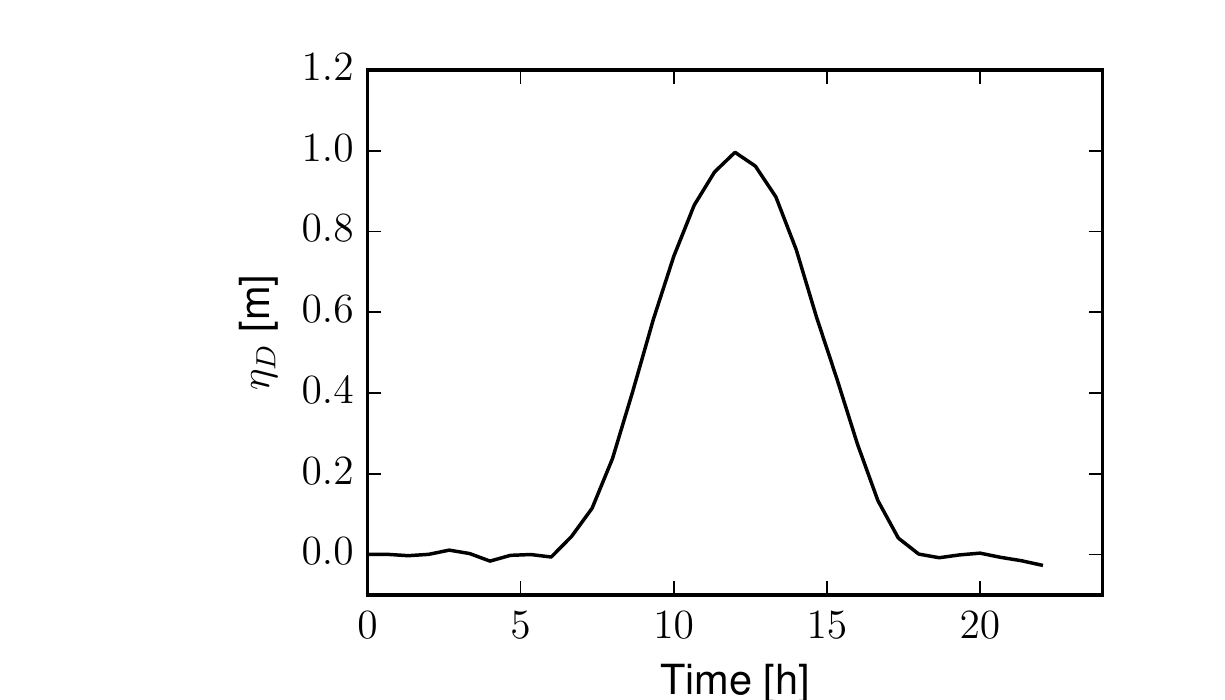}
            \caption*{Reconstructed wave, $\beta=4.0\cdot 10^{9}$}\label{fig:wetting_drying_balzano_0.43_noise_results_dg_beta_2000000000.0_noise_0.1_controls_final}
        \end{subfigure}
        \begin{subfigure}[b]{0.48\textwidth}
                \centering
		\includegraphics[width=1\textwidth]{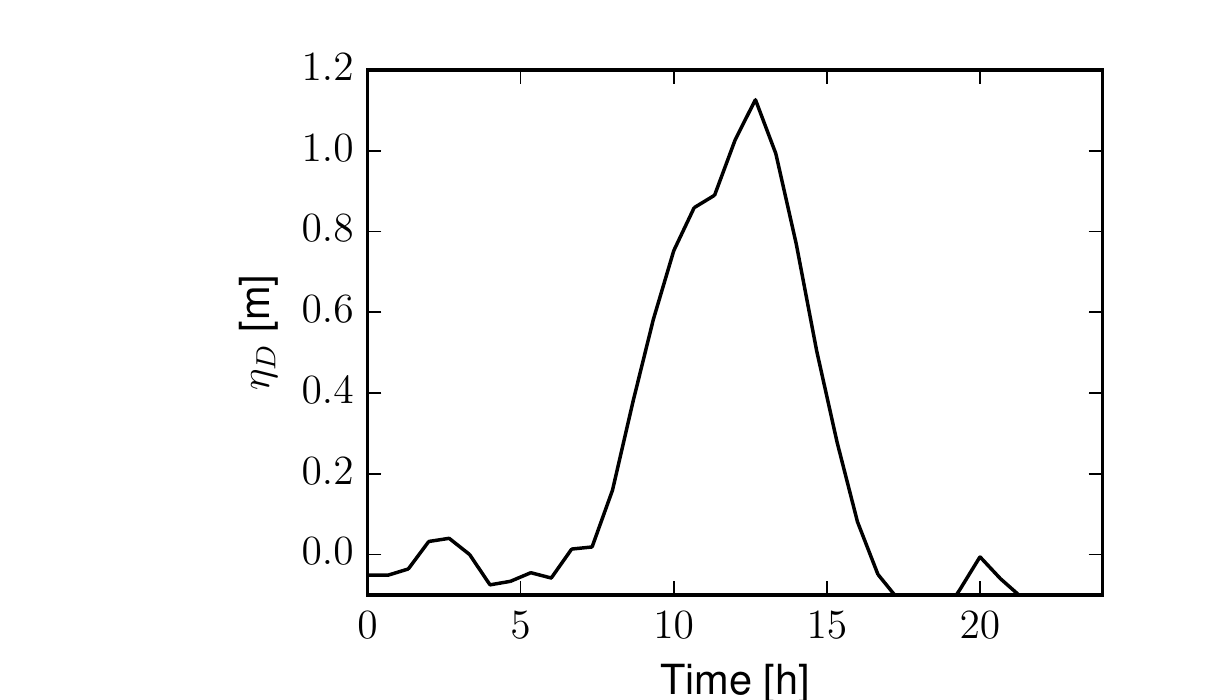}
            \caption*{Reconstructed wave, $\beta=4.0\cdot 10^{9}$}\label{fig:wetting_drying_balzano_0.43_noise_results_dg_beta_2000000000.0_noise_0.5_controls_final}
        \end{subfigure}\\
        \begin{subfigure}[b]{0.48\textwidth}
                \centering
		\includegraphics[width=1\textwidth]{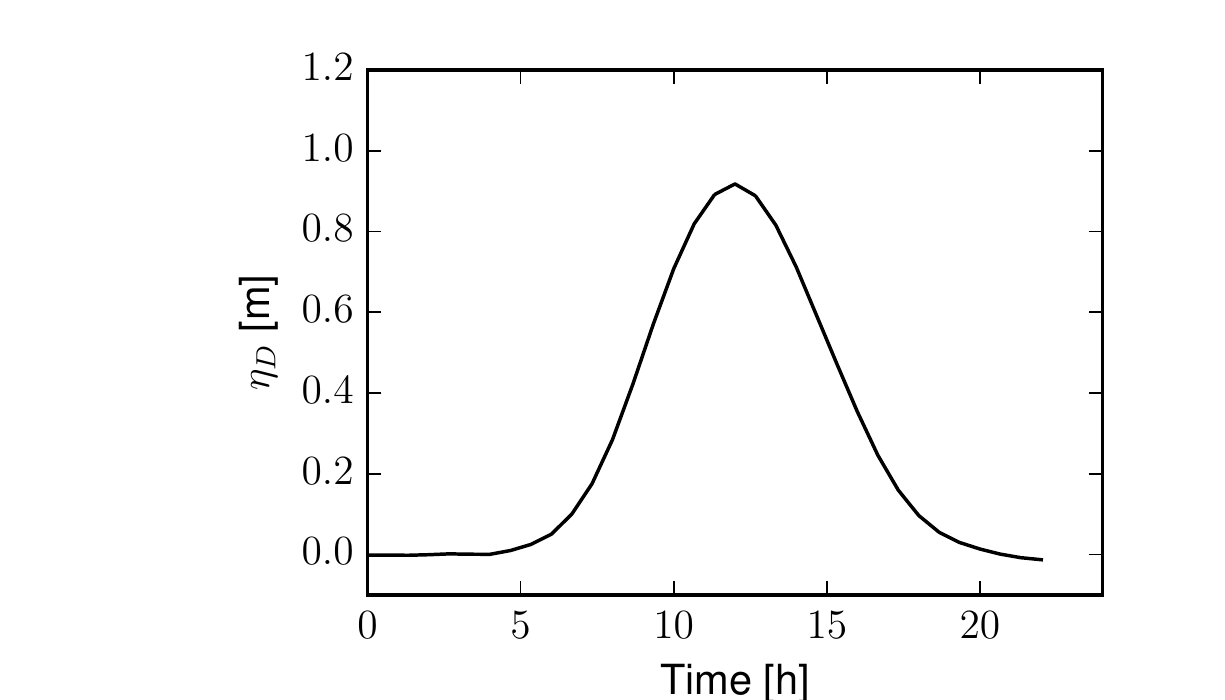}
            \caption*{Reconstructed wave, $\beta=4.0\cdot 10^{10}$}\label{fig:wetting_drying_balzano_0.43_noise_results_dg_beta_20000000000.0_noise_0.1_controls_final}
        \end{subfigure}
        \begin{subfigure}[b]{0.48\textwidth}
                \centering
		\includegraphics[width=1\textwidth]{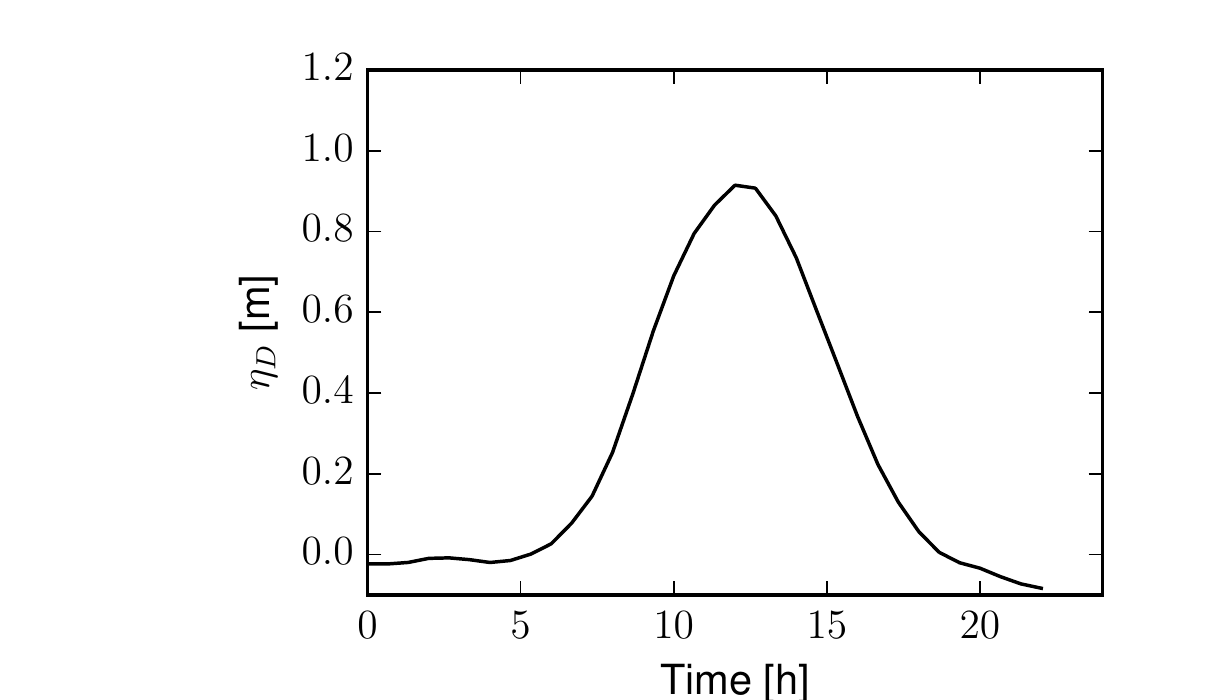}
            \caption*{Reconstructed wave, $\beta=4.0\cdot 10^{10}$}\label{fig:wetting_drying_balzano_0.43_noise_results_dg_beta_20000000000.0_noise_0.5_controls_final}
        \end{subfigure}\\
        \begin{subfigure}[b]{0.48\textwidth}
                \centering
		\includegraphics[width=1\textwidth]{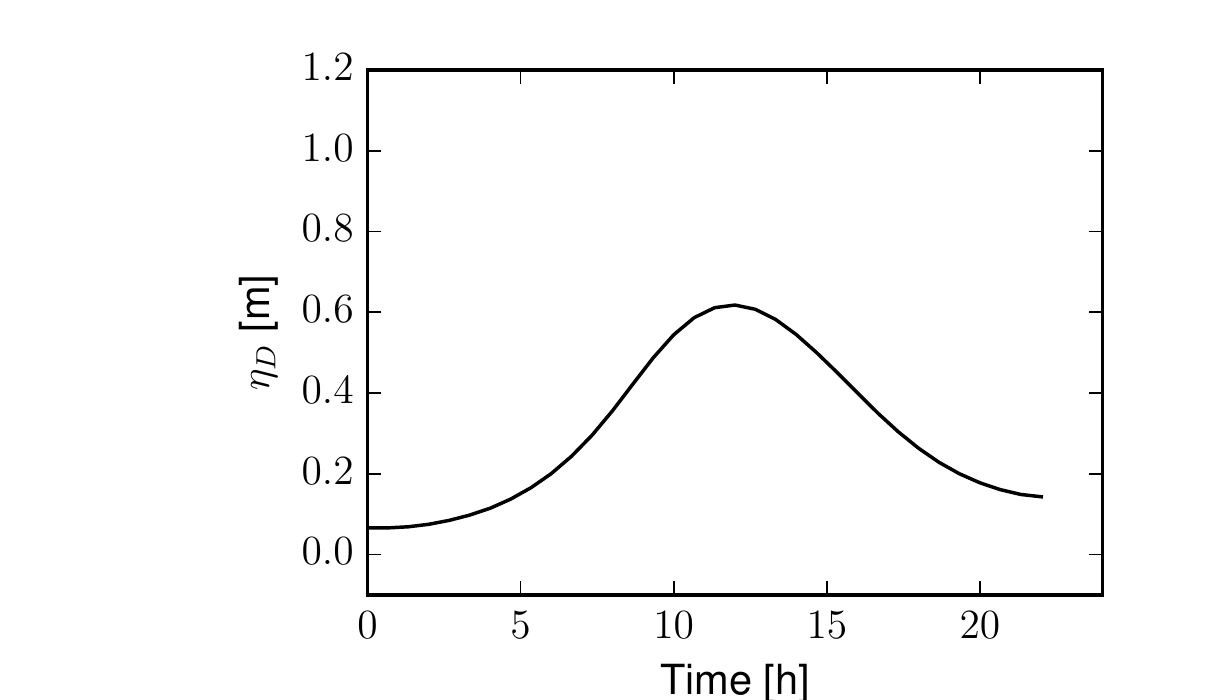}
            \caption*{Reconstructed wave, $\beta=4.0\cdot 10^{11}$}\label{fig:wetting_drying_balzano_0.43_noise_results_dg_beta_2e+11_noise_0.1_controls_final}
        \end{subfigure}
        \begin{subfigure}[b]{0.48\textwidth}
                \centering
		\includegraphics[width=1\textwidth]{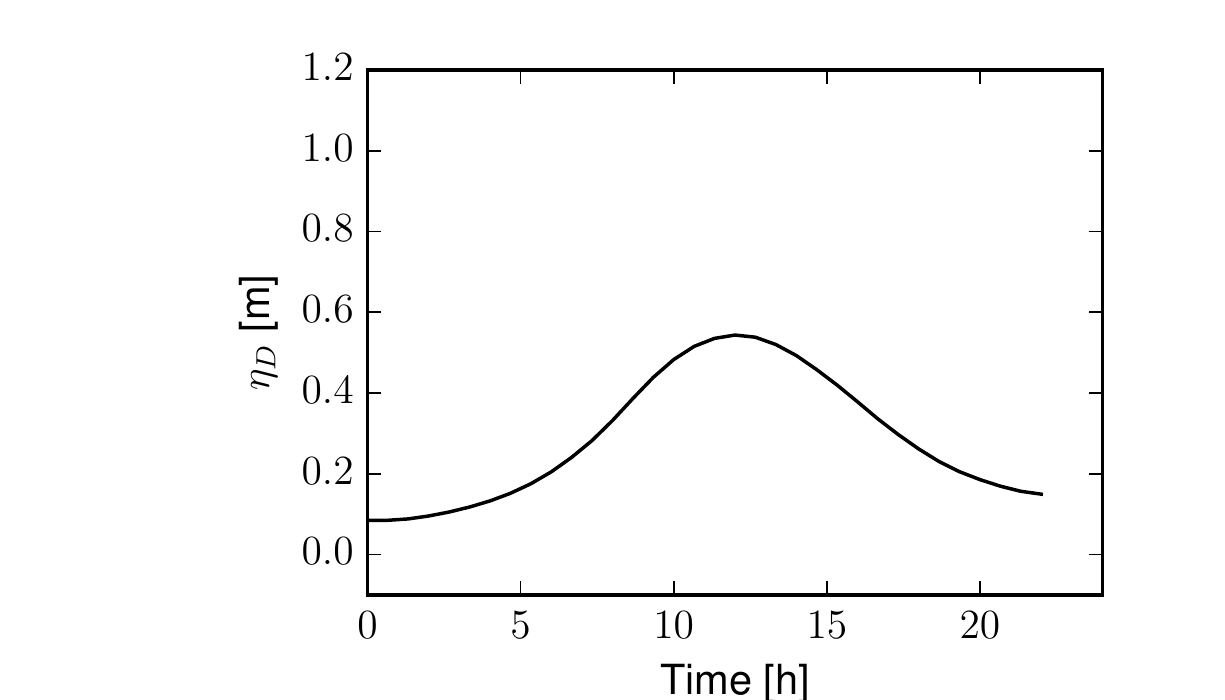}
            \caption*{Reconstructed wave, $\beta=4.0\cdot 10^{11}$}\label{fig:wetting_drying_balzano_0.43_noise_results_dg_beta_2e+11_noise_0.5_controls_final}
        \end{subfigure}
        \caption{Results of the wave profile reconstruction on a sloping beach with a sinusoidal incoming wave profile, a smoothing value $\alpha = 0.43$~m and pointwise Gaussian noise added to the observations with standard deviation $\sigma=0.1$ (left) and $\sigma=0.5$ (right). Different regularisation values $\beta$ were used to control the smoothness of the reconstructed wave. The final $2$ h of the wave profile are excluded from the reconstruction.}
\label{fig:wd_balzano1_results_alpha_0.43_noise/results_dg_noise_0.1_and_0.5}
\end{figure}

So far the experiments were performed with perfect observations in the
sense that the same model was used to produce the observations and to
reconstruct the wave profile. To avoid this `inverse crime', we repeated the experiment
with pointwise Gaussian noise added to the observations $d$. 
The noisy observations are shown on the upper frames of figure \ref{fig:wd_balzano1_results_alpha_0.43_noise/results_dg_noise_0.1_and_0.5} for two different noise levels. The impact of the noise on the observations is clearly visible. 

With noisy observations it becomes important to regularise the problem in order to avoid
the model describing the noisy data rather than the physical relationships, also known as overfitting.
For comparison, we repeated the reconstruction for three different regularisation values: $\beta = 4\times 10^9, 4\times 10^{10}, 4\times 10^{11}$. These were chosen such that the functional terms are approximately of the same magnitude at the beginning of the optimisation.
The results are shown in figure \ref{fig:wd_balzano1_results_alpha_0.43_noise/results_dg_noise_0.1_and_0.5}.
The quality of the reconstructed wave profiles depends clearly on the regularisation term. 
The larger $\beta$ is, the smoother and flatter the wave profile becomes. Considering 
the reference wave profile \ref{fig:wetting_drying_balzano_0.43_composed_sin_controls_optimal}, a value of $\beta=4 \times 10^{10}$ yields robust results for both noise levels in this example. 

\subsubsection{Composed sinusoidal wave profile}
\begin{figure}[p]
\centering
        \begin{subfigure}[b]{0.48\textwidth}
        \centering
		\includegraphics[width=\textwidth]{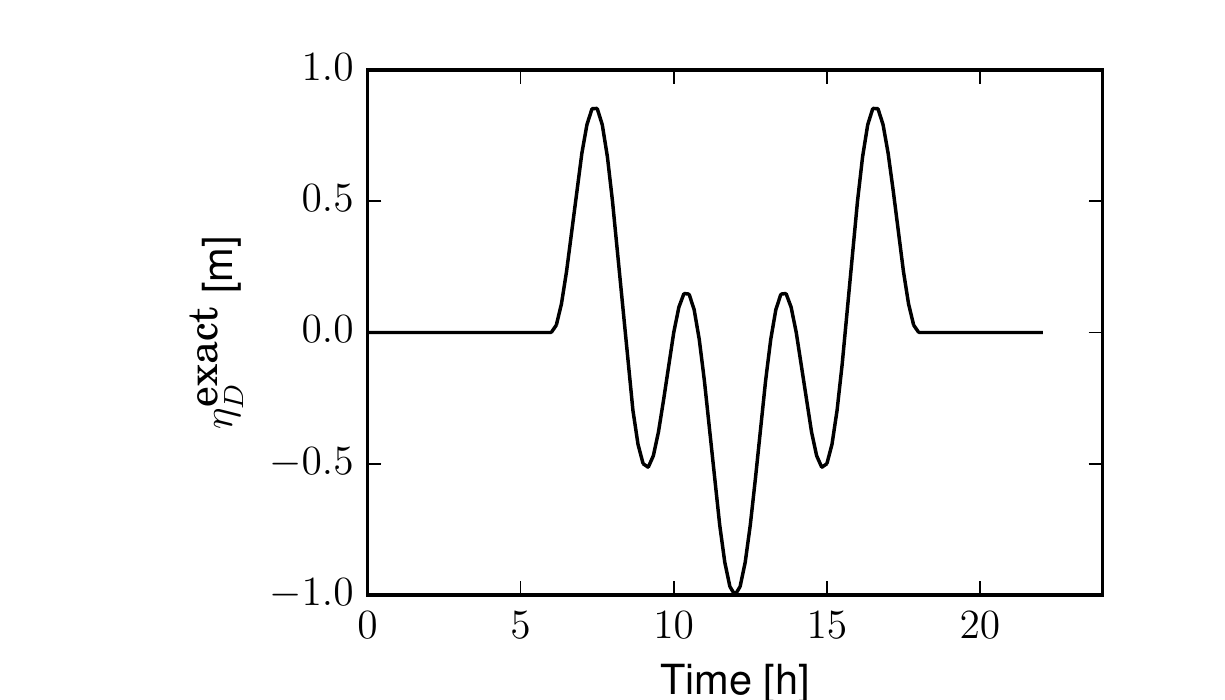}
		\caption{Reference Dirichlet wave profile\\~}\label{fig:wetting_drying_balzano_0.43_composed_sin_controls_optimal}
        \end{subfigure}
        \begin{subfigure}[b]{0.48\textwidth}
                \centering
		\includegraphics[width=\textwidth]{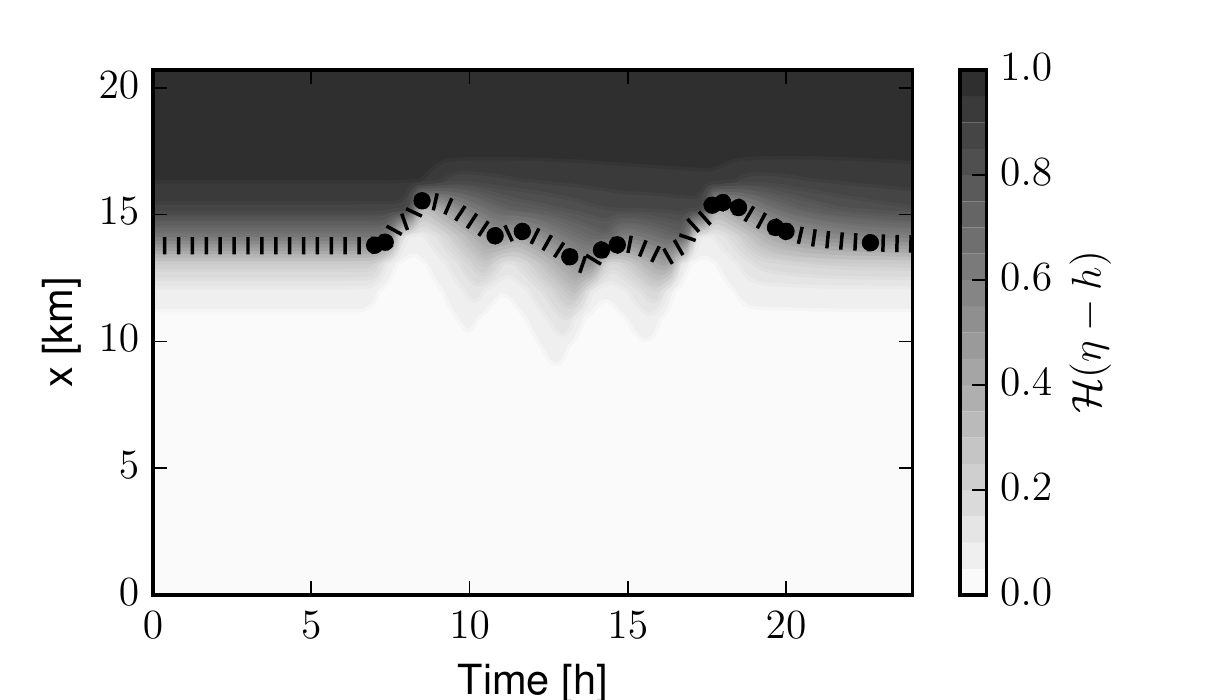}
		\caption{Observations and exact wet/dry interface (dashed line)}\label{fig:wetting_drying_balzano_0.43_plot_wd_interface_composed_sin}
        \end{subfigure}
        \\
        \begin{subfigure}[b]{0.48\textwidth}
                \centering
		\includegraphics[width=1\textwidth]{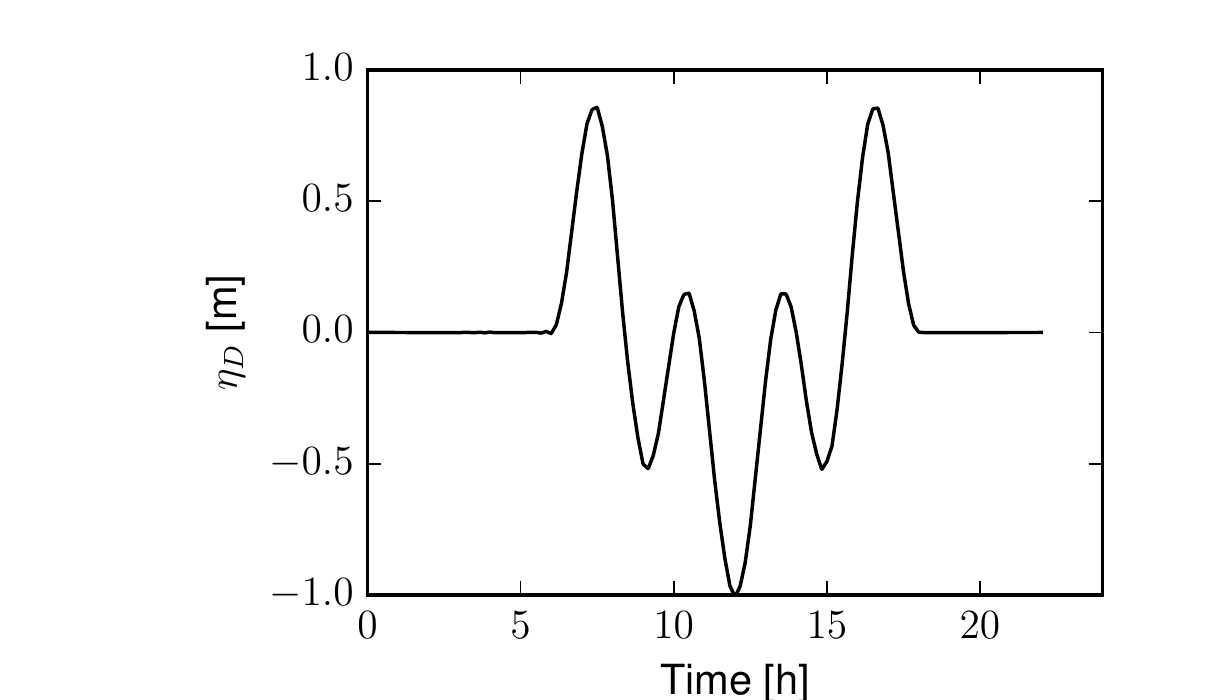}
		\caption{Reconstructed wave profile}\label{fig:wetting_drying_balzano_0.43_composed_sin_controls_101}
        \end{subfigure}
        \begin{subfigure}[b]{0.48\textwidth}
                \centering
		\includegraphics[width=1\textwidth]{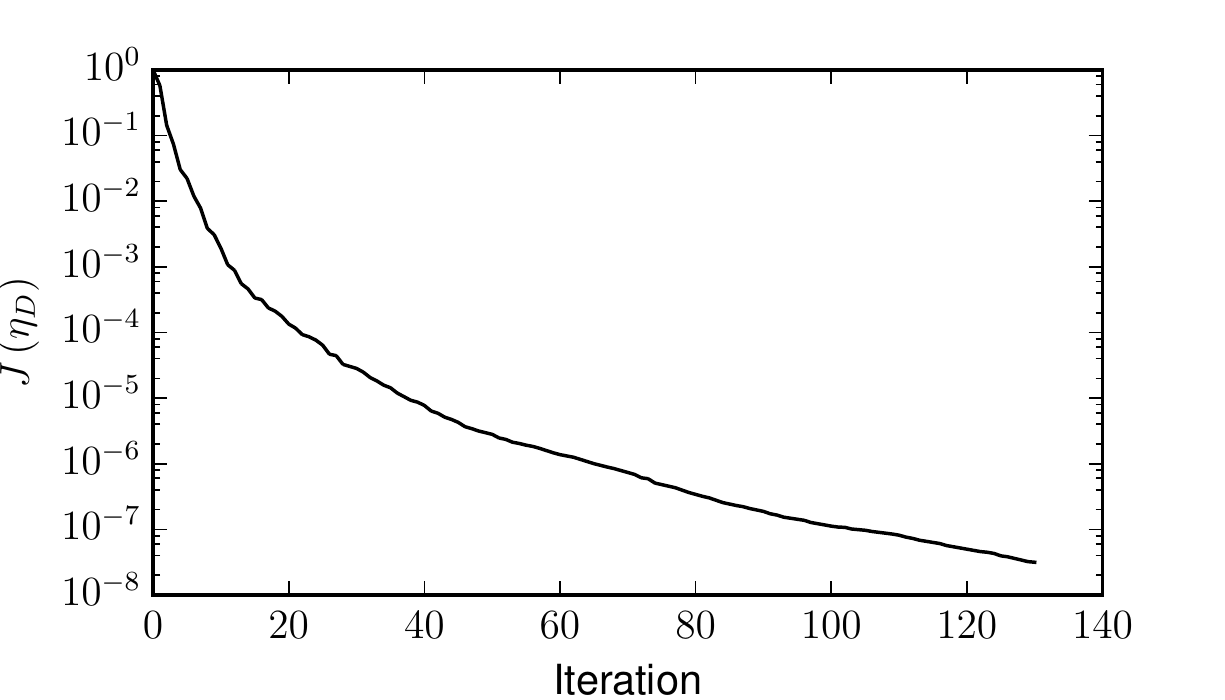}
		\caption{Optimisation convergence}\label{fig:wetting_drying_balzano_0.43_composed_sin_iter_plot}
        \end{subfigure}
        \caption{Results of the wave profile reconstruction on a sloping beach with a composed sinusoidal incoming wave profile and smoothing value $\alpha = 0.43$~m.
                 The final $2$ h of the wave profile are excluded from the reconstruction.}
\label{fig:wd_balzano1_results_alpha_0.43_composed}
\end{figure}

The next experiment demonstrates that a more complex wave profile can be reconstructed.
For that, the previous example is repeated with a reference Dirichlet function $\eta_D^{\textrm{exact}}$ that is the composition of two sinusoidal functions with different periods (figure~\ref{fig:wetting_drying_balzano_0.43_composed_sin_controls_optimal}):
\begin{equation*}
\eta_{D}^{\textrm{exact}}(t) =
\begin{cases}
  0 & \textrm{if } t < 6 \textrm{ h}, \\
\frac{1}{2}\left(\cos\left(\frac{2\pi (t - 6 \textrm{ h})}{p}\right) - \cos\left(\frac{8 \pi (t- 6 \textrm{ h})}{p}\right)\right) & \textrm{if } 6 \textrm{ h} \le t \le 18 \textrm{ h}, \\
0 & t > 18 \textrm{ h.}
\end{cases}
\end{equation*}
where $p\coloneqq 12$~h.
The smoothing value is again $\alpha=0.43$~m.
The observations are generated in the same way as in example \ref{sec:sin_no_noise} and are plotted in figure~\ref{fig:wetting_drying_balzano_0.43_plot_wd_interface_composed_sin}.

In this case, the optimisation tolerance was reached after $141$ iterations.
The results in figure~\ref{fig:wd_balzano1_results_alpha_0.43_composed} show that the wave was successfully reconstructed. The maximum error is $1.2$ cm, or a relative error of $1.2$ \%.
Comparing the required numbers of iterations to the previous experiments indicates that the shape/complexity of the wave profile to be reconstructed impacts the convergence rate of the optimisation method.

\subsection{Reconstruction of the Hokkaido-Nansei-Oki tsunami wave profile}
\begin{figure}[bt]
\centering
  \begin{subfigure}[b]{0.48\textwidth}
      \centering
      \includegraphics[width=\textwidth]{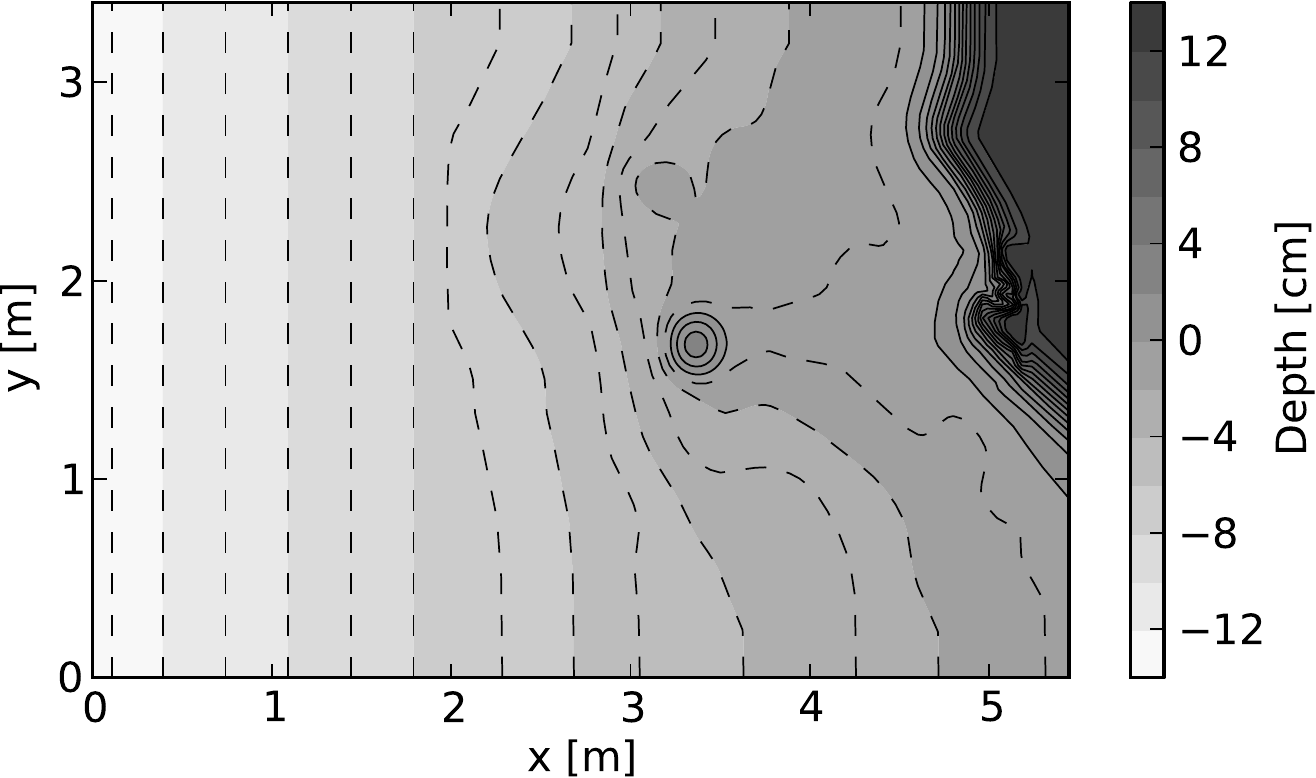}
      \caption{Bathymetry}\label{fig:wd_optimal_rate_optimal_monai_valley_bathymetry}
  \end{subfigure}
  \begin{subfigure}[b]{0.48\textwidth}
      \centering
      \includegraphics[width=0.86\textwidth]{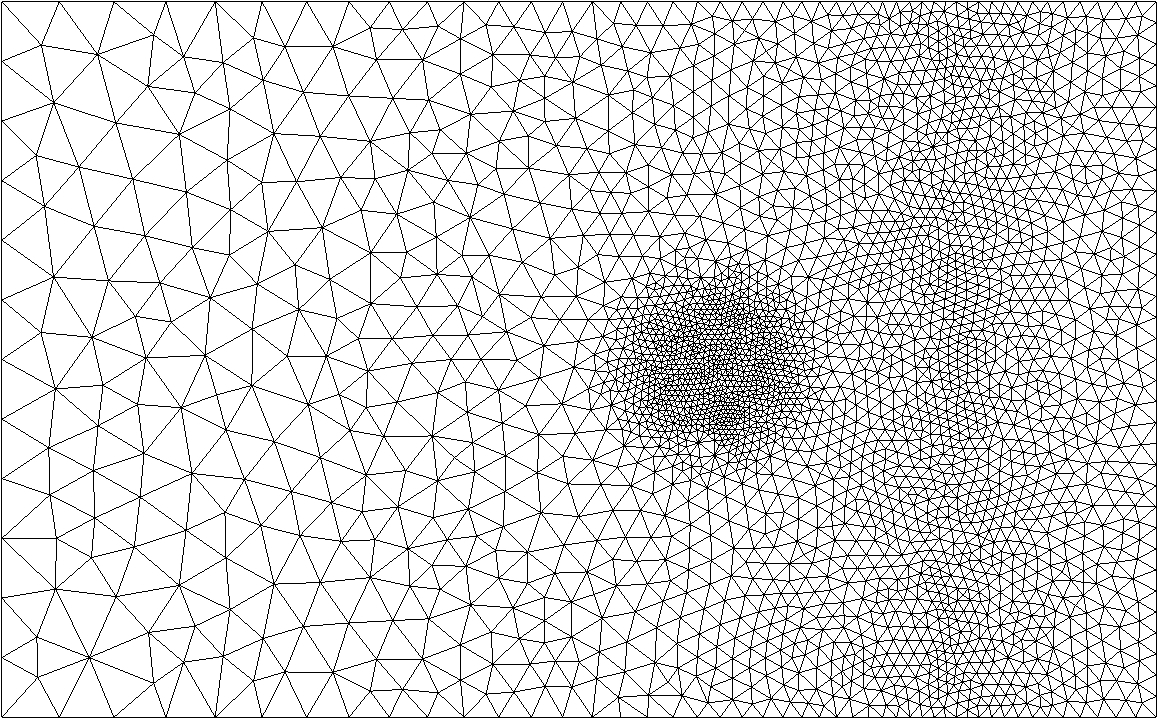}
      \vspace{0.45cm}
      \caption{Mesh}\label{fig:wd_optimal_rate_optimal_monai_valley_mesh}
  \end{subfigure}
  \caption{The laboratory setup of the Hokkaido-Nansei-Oki tsunami example, based on $1/400$ laboratory experiment by the Central Research Institute for Electric Power Industry. The island at the centre and the coast on the right are hit by a tsunami shaped wave coming from the left boundary.}
\end{figure}

\begin{figure}[p]
\centering
        \begin{subfigure}[b]{\textwidth}
                \centering
		\includegraphics[width=0.4\textwidth]{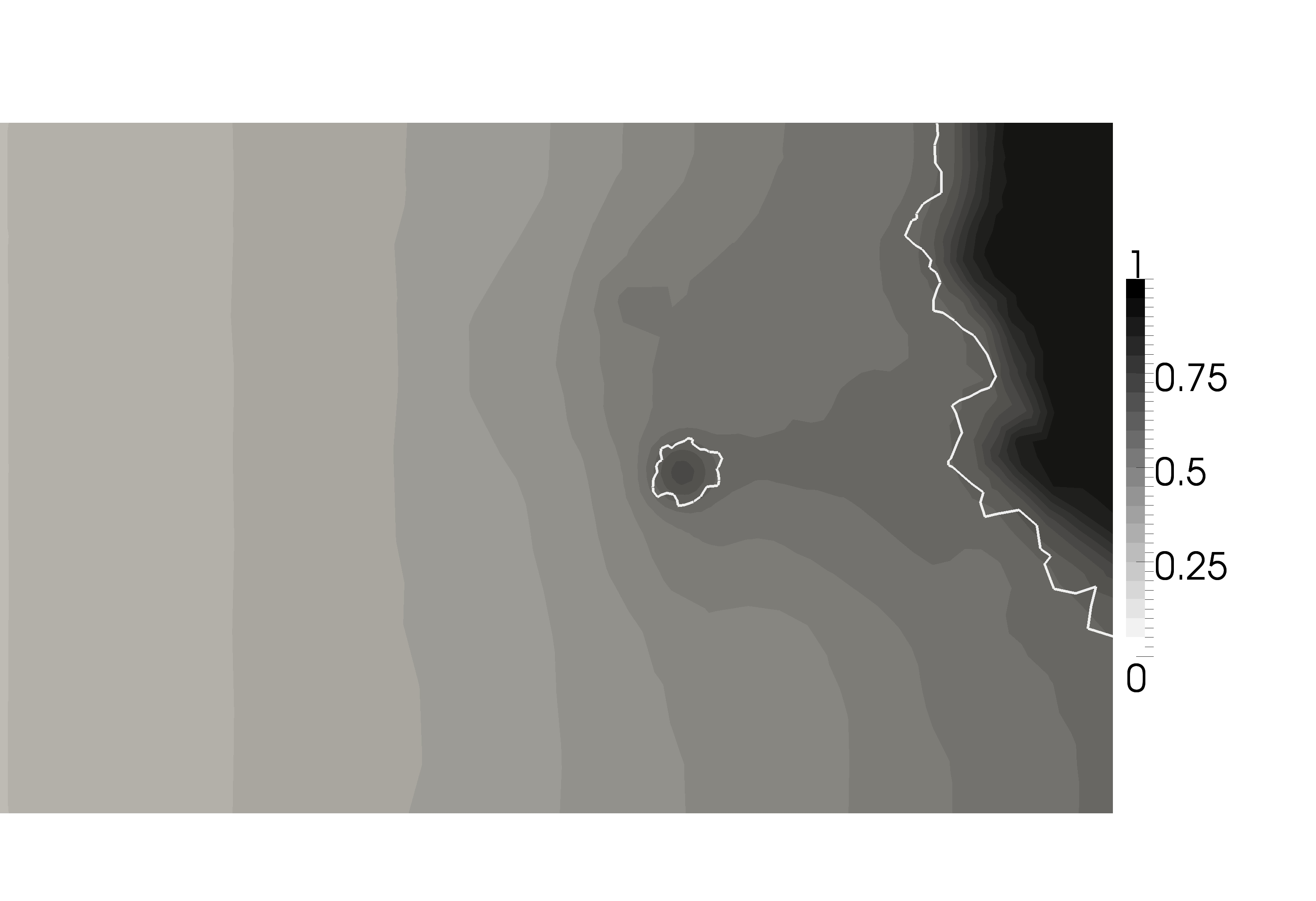}
		\includegraphics[width=0.4\textwidth]{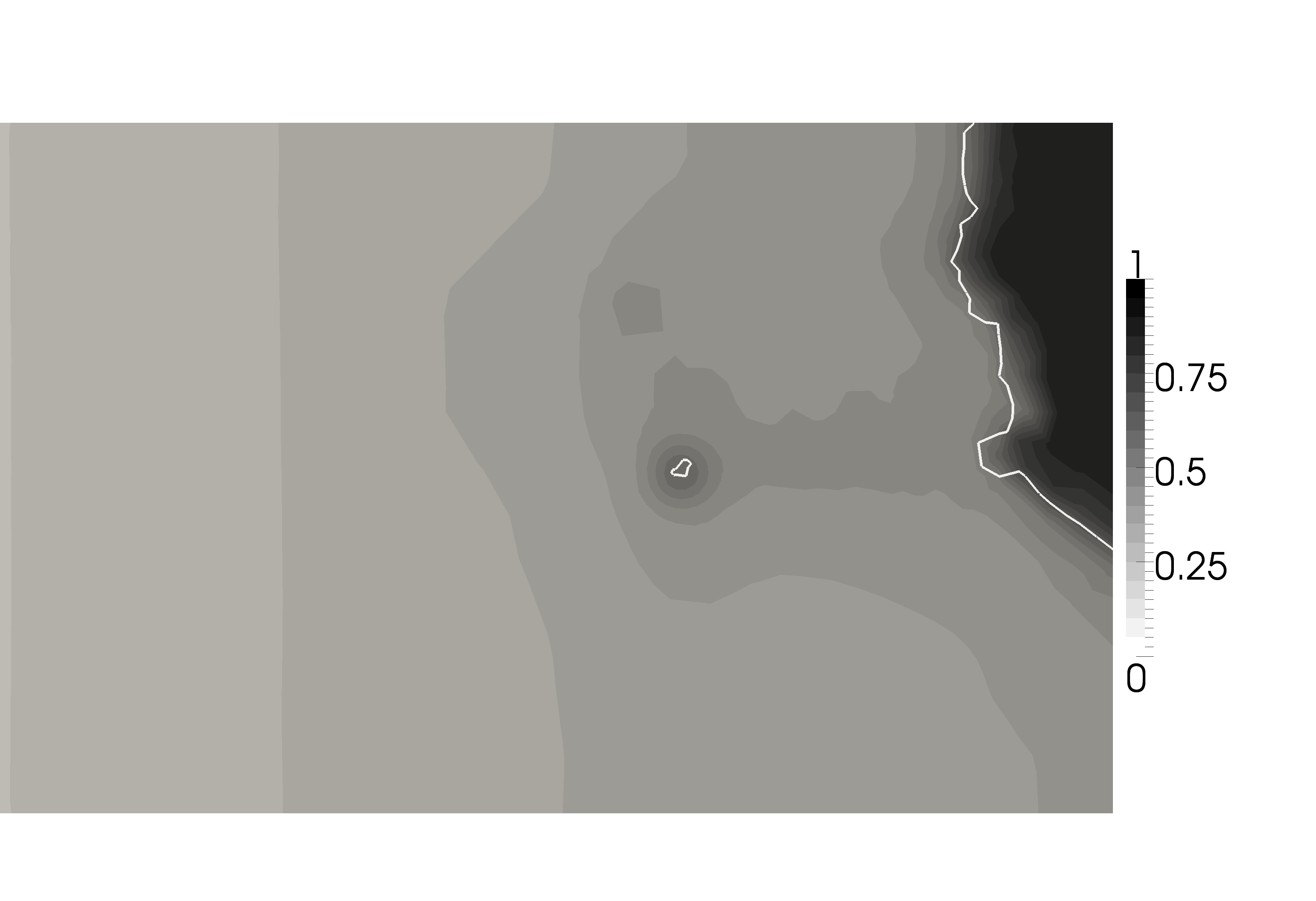}
		\includegraphics[width=0.4\textwidth]{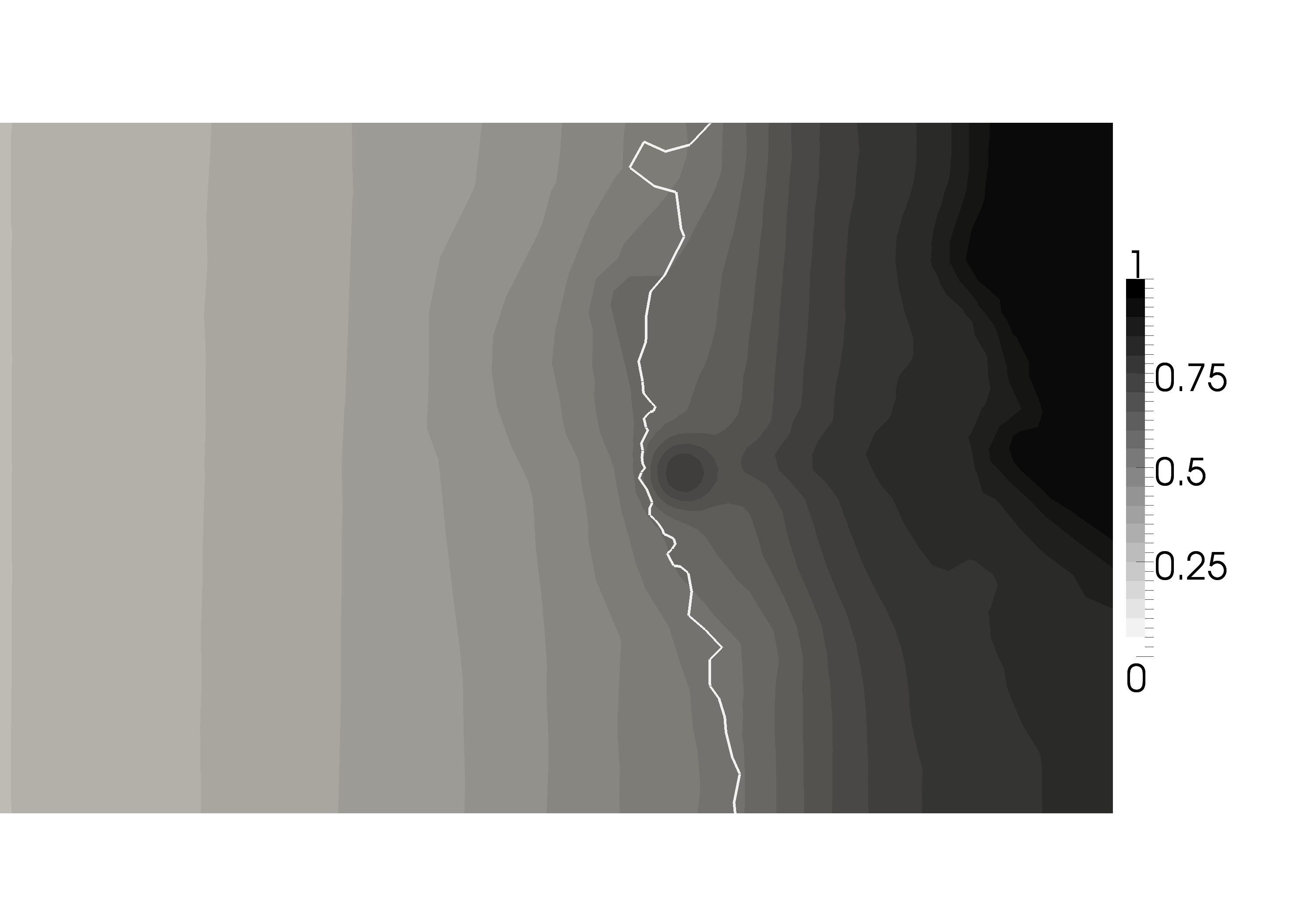}
		\includegraphics[width=0.4\textwidth]{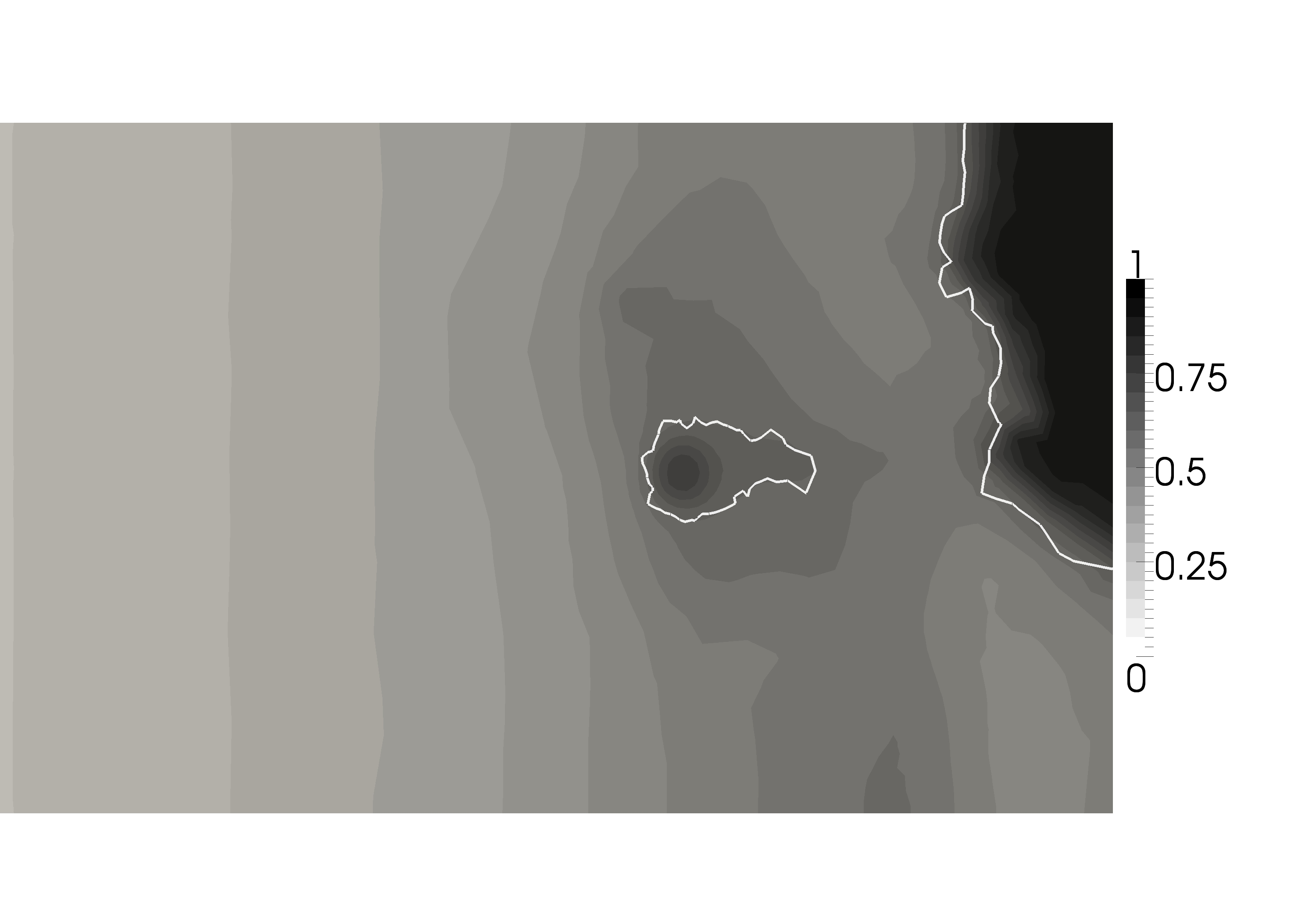}
		\caption{The wet/dry interface observations after $0$ s, $20$ s, $29$ s, $32$ s, in reading order. The observations are constructed by running the forward problem with the synthetic Dirichlet boundary values.
			The observations are approximated indicator functions of the wet/dry interface (marked as white lines)
}\label{fig:wetting_drying_tsunami_observations}
        \end{subfigure}
        \\
        \begin{subfigure}[b]{0.48\textwidth}
                \centering
		\includegraphics[width=1\textwidth]{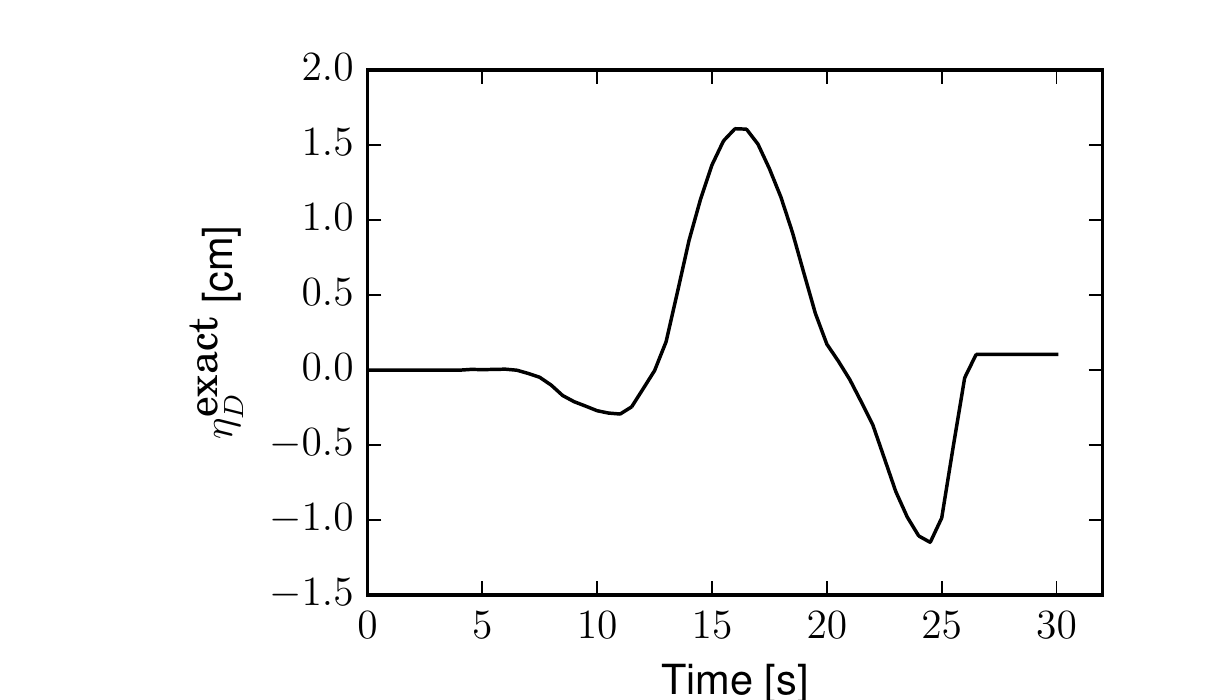}
		\caption{The reference tsunami wave profile.\\~}\label{fig:hokkaido-nansei-oki_tsunami_controls_optimal}
        \end{subfigure}
        \begin{subfigure}[b]{0.48\textwidth}
                \centering
		\includegraphics[width=1\textwidth]{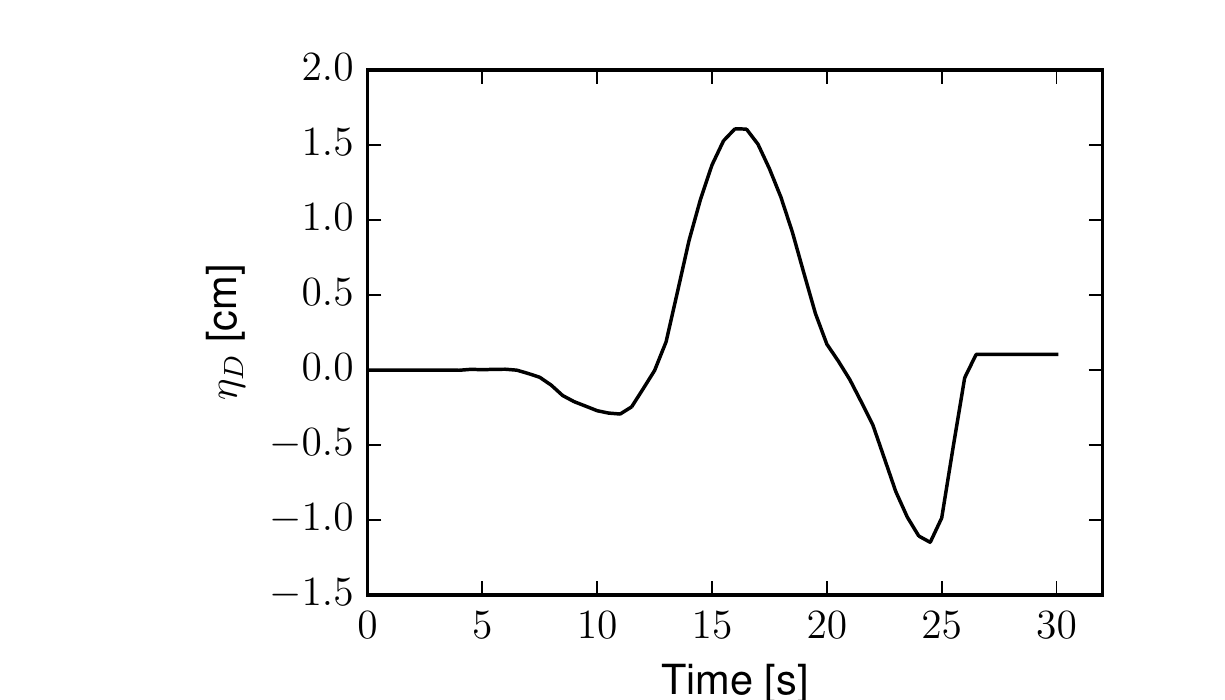}
		\caption{The reconstructed tsunami wave profile.}\label{fig:wetting_drying_hokkaido-nansei_tsunami_controls_final}
        \end{subfigure}
        ~
        \begin{subfigure}[b]{0.48\textwidth}
                \centering
		\includegraphics[width=1\textwidth]{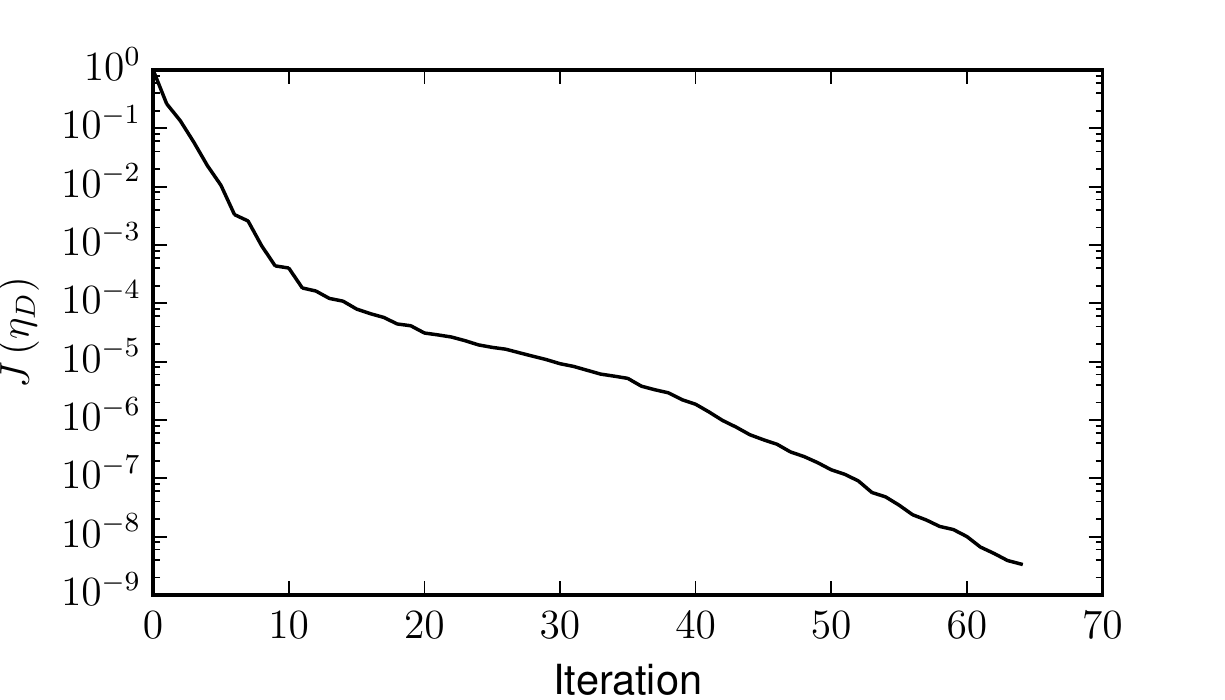}
		\caption{The functional values during the optimisation iterations\\~}\label{fig:wetting_drying_tsunami_iter_plot}
        \end{subfigure}
        \caption{Results of the reconstruction of the Hokkaido-Nansei-Oki tsunami wave profile.
                 The final $2$ s of the boundary values are excluded from the reconstruction.}
\label{fig:hokkaido-nansei-oki_tsunami-results}
\end{figure}

The second data assimilation problem is motivated by the question of whether it is possible to reconstruct a tsunami wave profile from satellite observations that record the inundation line over time.
The considered event is the Hokkaido-Nansei-Oki tsunami that occurred in 1993 and produced run-up heights of up to $30$~m on Okushiri island, Japan.
The Central Research Institute for Electric Power Industry (CRIEPI) in Abiko, Japan constructed a $1/400$ scale laboratory model of the area around the island \citep{matsuyama2001}.
Following~\citet{yalciner2011}, this experiment is simulated in a rectangular domain of size $5.448\textrm{ m} \times 3.402\textrm{ m}$.
The bathymetry and coastal topography is shown in figure \ref{fig:wd_optimal_rate_optimal_monai_valley_bathymetry}.
It contains an island in the centre and coastal regions on the top right of the domain.
On the left boundary a surface elevation profile is enforced that resembles a tsunami wave (figure \ref{fig:hokkaido-nansei-oki_tsunami_controls_optimal}).
The aim of this experiment is to reconstruct this wave profile.
On the remaining boundaries, a no-normal flow condition is imposed.

The domain is discretised with an unstructured mesh consisting of $5,730$ triangular elements with increasing resolution near the inundation areas, see figure~\ref{fig:wd_optimal_rate_optimal_monai_valley_mesh}. 
The mesh elements size range from $0.4$ m to $0.02$ m.
The temporal discretisation uses a time step of $0.5$~s with a total simulation time of $32$~s. The Manning coefficient is set to $\mu=0.025\textrm{ s}/\textrm{m}^{\frac{1}{3}}$.
The observations are synthetically generated by running the shallow water model with the reference wave profile used in the laboratory experiment while recording the wet/dry interface. No noise was added, and $\beta=0$ used.
The smoothing guideline equation~\eqref{eq:epsilon_definition_in_wd} yields $\alpha\approx0.16$~m, in the experiments we chose a value of $\alpha =0.1$~m.


The optimisation was initialised with a wave profile of $1.05\times 10^{-3}$ m for all time levels, which corresponds to the final free-surface displacement of the input wave.
For the same reason as in the examples above, the final $2$ s of the Dirichlet boundary values were then reset to the reference Dirichlet boundary values and excluded from the reconstruction.
Furthermore, a box constraint was used to restrict the minimum and maximum free-surface displacement to $-1.5$ cm and $+2$ cm.
Without these constraints the optimisation generated unrealistically large Dirichlet boundary values at an intermediate iteration for which the Newton solver in the shallow water model diverged.

The optimisation iteration converged after $67$ iterations.
The results are shown in figure~\ref{fig:hokkaido-nansei-oki_tsunami-results}.
The incoming wave was reconstructed up to an absolute error of $9 \times 10^{-4}$ cm,
or a relative error of less than $6\times 10^{-4}$\%.

\section{Summary}
A shallow water model with wetting and drying and its adjoint model has been
developed and used to reconstruct an incoming wave profile from inundation
observations. The reconstruction is formulated as an optimisation problem
which minimises the difference between the observed and the simulated wet/dry
interface and solved with an efficient gradient-based optimisation method.
This problem setup is a step towards reconstructing unknowns such as the tsunami source and wave profile from real inundation data that is available from historical data or satellite imaging.

Numerical experiments demonstrate that, under idealised conditions, the profile
of the incoming wave can be accurately recovered. Furthermore, an experiment with added Gaussian noise in the observations showed robustness with respect to noisy data. 
However, multiple question remain unanswered. In particular, we lack convergence analysis for the regularised continuous problem to the nonsmooth, free-boundary problem. Similarily, convergence analysis of the discretisation of the regularised continuous problem is lacking. Furthermore, a mesh-independent optimisation method should be employed to obtain improve performance, in particular for setups with spatially varying wave profiles or adaptive-timestepping. Finally, experiments with real observations are needed to fully exclude inverse crimes.

The initial results of the paper are promising and motivate future research in
this direction: for example, the robustness of the data assimilation should be
tested against partially missing observations. In this case,
the regularisation parameter $\beta$ will play an important role to enforce smoothness of the reconstructed wave profile. Another direction is to overcome the shallow water assumption, as it might not accurately capture important physical processes. One possibility is to replace the shallow water model with a three-dimensional wetting and drying model.

\section*{Code availability}
The model implementation and files needed to reproduce the results of this paper are freely available on bitbucket:
\url{https://bitbucket.org/simon_funke/wetting_and_drying_optimisation_code}.
The website contains a Readme file with instructions for
how to install the software and reproduce the results of the paper.

\section*{Acknowledgements}
This work was supported by the Grantham Institute for Climate Change, the Research Council of Norway through a Centres of Excellence grant to the Center for Biomedical Computing at Simula Research Laboratory (project number 179578), a FRIPRO grant (project number 251237), a UK Engineering and Physical Sciences Research Council grant (EP/K030930/1) and a UK Natural Environment Research Council grant (NE/K000047/1).


\bibliographystyle{elsarticle-harv}
\bibliography{literature}

\end{document}